\documentclass[twoside,twocolumn,english,aps,showpacs,prx,reprint,amsmath,amssymb,notitlepage,superscriptaddress,longbibliography]{revtex4-1}
\usepackage[T1]{fontenc}
\usepackage[latin9]{inputenc}
\usepackage{color}
\usepackage{bm}
\usepackage{subfigure}
\usepackage{amsmath}
\usepackage{amssymb}

\usepackage{graphicx}
\usepackage[left=2cm, right=2cm, top=2cm, bottom=2cm]{geometry}

\usepackage{esint}
\usepackage{mathrsfs}
\usepackage{booktabs}
\usepackage{multirow}
\usepackage{makecell}
\usepackage{braket}
\usepackage{babel}
\usepackage{extarrows}

\usepackage{color}
\usepackage{longtable}
\usepackage{verbatim}
\usepackage{natbib}
\usepackage{blindtext}
\usepackage[titletoc,title]{appendix} 
\usepackage{titlesec} 

\usepackage[plainpages = false, pdfpagelabels, 
                 bookmarks,
                 bookmarksopen = true,
                 bookmarksnumbered = true,
                 breaklinks = true,
                 linktocpage,
                 colorlinks = true,
                 linkcolor = blue,
                 urlcolor  = blue,
                 citecolor = blue,
                 anchorcolor = green,
                 hyperindex = true,
                 hyperfigures
                 ]{hyperref} 
\usepackage{dcolumn}   % needed for some tables

\makeatletter

\makeatletter
\newcommand*{\rom}[1]{\expandafter\@slowromancap\romannumeral #1@}
\makeatother

\usepackage{times}{\Large }

\renewcommand{\selectlanguage}[1]{}

	\DeclareMathOperator{\tr}{tr}  		% matrix trace
	\DeclareMathAlphabet{\mathbbold}{U}{bbold}{m}{n}
        \def\ii{\mathrm{i}}
        \def\e{\mathrm{e}}

\makeatother

\begin{document}

\title{Vulnerability of fault-tolerant topological quantum error correction to quantum deviations in code space}

\author{Yuanchen Zhao}
\affiliation{State Key Laboratory of Low Dimensional Quantum Physics, Department of Physics, Tsinghua University, Beijing, 100084, China}
\affiliation{Frontier Science Center for Quantum Information, Beijing 100184, China}

\author{Dong E. Liu}
\email{Corresponding to: dongeliu@mail.tsinghua.edu.cn}
\affiliation{State Key Laboratory of Low Dimensional Quantum Physics, Department of Physics, Tsinghua University, Beijing, 100084, China}
\affiliation{Frontier Science Center for Quantum Information, Beijing 100184, China}
\affiliation{Beijing Academy of Quantum Information Sciences, Beijing 100193, China}
\affiliation{Hefei National Laboratory, Hefei 230088, China}
%\affiliation{Beijing Academy of Quantum Information Sciences, Beijing 100193, China}

\begin{abstract}
Quantum computers face significant challenges from quantum deviations or coherent noise, particularly during gate operations, which pose a complex threat to the efficacy of quantum error correction (QEC) protocols. In this study, we scrutinize the performance of the topological toric code in 2 dimension (2D) under the dual influence of stochastic noise and quantum deviations, especially during the critical phases of initial state preparation and error detection facilitated by multi-qubit entanglement gates.  By mapping the protocol for multi-round error detection--from the inception of an imperfectly prepared code state via imperfect stabilizer measurements--to a statistical mechanical model characterized by a 3-dimensional $\mathbb{Z}_2$ gauge theory coupled with a 2-dimensional $\mathbb{Z}_2$ gauge theory, we establish a novel link between the error threshold and the model's phase transition point. Specifically, we find two distinct error thresholds that demarcate varying efficacies in error correction. The empirical threshold that signifies the operational success of QEC aligns with the theoretical ideal of flawless state preparation operations. Contrarily, below another finite theoretical threshold, a phenomenon absent in purely stochastic error models emerges: unidentifiable measurement errors precipitate QEC failure in scenarios with large code distances. For codes of finite or modest distance $d$, it is revealed that maintaining the preparation error rate beneath a crossover scale, proportional to $1/\log d$, allows for the suppression of logical errors.  Considering that fault-tolerant quantum computation is valuable only in systems with large scale and exceptionally low logical error rates, this investigation explicitly demonstrates the serious vulnerability of fault tolerant QEC based on 2D toric codes to quantum deviations in code space, highlighting the imperative to address inherent preparation noise.
\end{abstract}


\date{\today}

\maketitle

%\tableofcontents

\section{Introduction}
Quantum supremacy was recently observed in  quantum processors~\cite{arute,Zhong-USTC2020-Science,wu2021strong}, which is a milestone in the field of quantum computation. 
%Because to their noisy character, 
However, the state-of-the-art quantum devices~\cite{arute,Zhong-USTC2020-Science,Arute-Google2020Science,Gong-USTC2021-Science,Erhard21Nature,Pino-TrapIon2021-Science,wu2021strong,ryananderson2021} are classified as noisy intermediate-scale quantum (NISQ)~\cite{Preskill2018NISQ} computer, and the observed quantum supremacy is only a weakened version with few practical applications~\cite{Preskill2018NISQ}. To date, the merely known examples with worthwhile quantum advantages are only expected in fault tolerant quantum computers with quantum error correction (QEC)~\cite{shor,stean,Calderbank96}. Recently, QEC codes with small size have recently been tested in experiments~\cite{schindlerExperimentalRepetitiveQuantum2011,NiggScience14,Ofek16Nature,Hu19NP,Andersen20NP,Erhard21Nature,GoogleAI21Nature,Luo21PNAS,Marques22NP,ZhaoPRL-22,Ryan-Anderson21PRX,Egan20arXiv,sundaresanDemonstratingMultiroundSubsystem2023,bluvstein_logical_2024}.  

%A key concept of fault tolerance is the ``error threshold theorem'', which states that if the physical error rates \textcolor{red}{for all steps during quantum computation} are below an error threshold, quantum computation with arbitrary logical accuracy can be implemented in the noisy quantum devices
A cornerstone of fault tolerance is the "error threshold theorem." It posits that if the physical error rates across all facets of quantum computation--including code state preparation, stabilizer checks, logical operations, and readout--remain beneath a finite threshold, then one can achieve arbitrary logical accuracy within a noisy quantum device~\cite{knill2,aharonov2,aliferis}.
The threshold theorem is well-established if the device noise can be captured by independent stochastic errors~\cite{knill2,ref:dennis,aliferis,fowler,kovalev_fault_2013,gottesman_fault-tolerant_2014,Bombin-arxiv,vuillot_quantum_2019}, including circuit-level noise models~\cite{Baireuther_2019,pryadko_maximum-likelihood_2020}. However, actual quantum devices suffer from more general type of errors.
%such as correlated errors and coherent errors due to non-Markovian environments, miscalibration and imperfect control of gate operations. 
With correlated errors, the threshold theorem is modified for a more conceptual infidelity measure, e.g. diamond norm~\cite{aharonov2}, for correlations with weak amplitude~\cite{aharonov} and short length~\cite{chubb}, and for the environment with critical behaviors~\cite{novais,novais2}. A more practical type of noise comes from imperfect calibration and control of gate operations, causing quantum deviation or coherent effect in errors. This problem motivated recent studies of the independent single-qubit coherent errors ~\cite{barnes,beale,bravyi,iversonCoherenceLogicalQuantum2019,ehuang,cai,ouyang,Zhao21arxiv,Venn} and the detection induced coherent errors from entanglement gate noise~\cite{debroyStabilizerSlicingCoherent2018,ref:yang,ref:guoyi}. We emphasize that the two-qubit entanglement gates are much harder to calibrate and more error-prone than single-qubit gates. 
Comprehending the fault-tolerance and error threshold theorem in practical quantum systems exhibiting coherent deviations proves to be formidable, primarily due to the scarcity of analytical and numerical methodologies.
%Nevertheless, the fault-tolerance and error threshold theorem is not established in practical quantum devices with general noise. Understanding this issue is exceedingly challenging due to the lack of analytical and numerical tools. 
Although an efficient numerical strategy exists for a special case~\cite{bravyi}, the general coherent error problems, that go beyond the Clifford algebra, cannot be simulated efficiently in classical computers. 
Additionally, previous studies on QEC error thresholds have seldom addressed quantum deviations in code state preparation and error detection.
This motivates us to develop a theoretical framework to study the elusive and unavoidable quantum deviations alone with stochastic errors in realistic QEC procedures and establish a more practical error threshold theorem.
%Of that situation, there is no solid foundation for fault-tolerant quantum computation with practical quantum devices. 
%This motivates us to build a theoretical framework to study the coherent error problems in QEC and threshold theorem.

\begin{figure}[t] 
    \includegraphics[width=1\columnwidth]{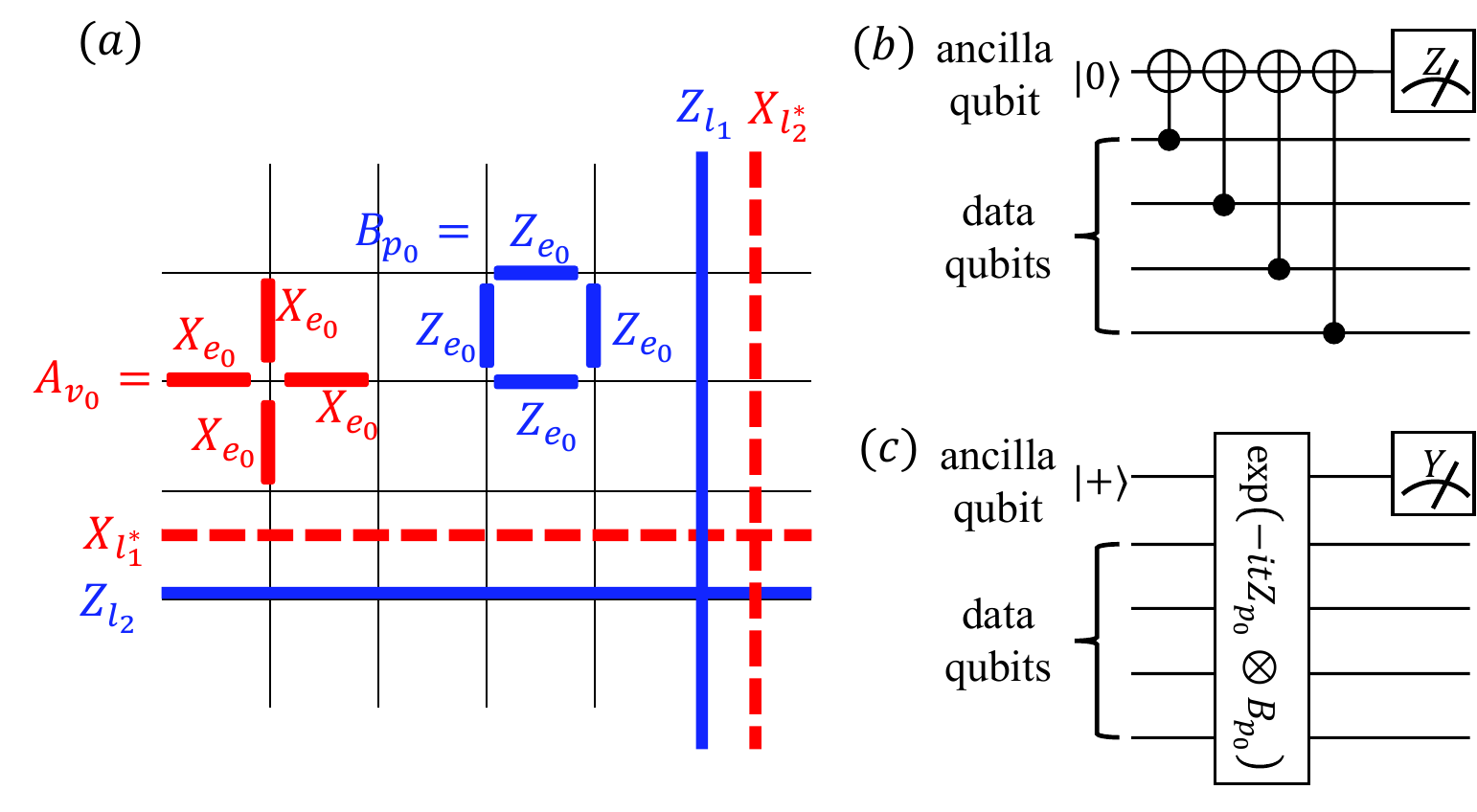}
    \caption{(a) Toric code defined on 2D periodic lattice. Physical qubits stay on the edges of the lattice. The two kinds of stabilizers are $A_{v_0}=\prod_{e_0|v_0\in\partial e_0}X_{e_0}$ defined on each vertex and $B_{p_0} = \prod_{e_0\in \partial p_0} Z_{e_0}$ defined on each plaquette as shown in the figure. 
    $l_1$ and $l_2$ denote non-contractible loops on the periodic lattice. The logical Pauli $Z$ operators $Z_{l_1}$ and $Z_{l_2}$ are product of $Z$'s along these non-contractible loops. Correspondingly the logical Pauli $X$ operators $X_{l_1^*}$ and $X_{l_2^*}$ are defined as $X$'s along non-contractible loops on the dual lattice $l_1^*$ and $l_2^*$.
    They satisfy the commutation relations $X_{l^*_1}Z_{l_1}=-Z_{l_1}X_{l^*_1}$, $X_{l^*_1}Z_{l_1}=-Z_{l_1}X_{l^*_1}$, $X_{l^*_1}Z_{l_2}=Z_{l_2}X_{l^*_1}$, $X_{l^*_2}Z_{l_1}=Z_{l_1}X_{l^*_2}$ and acts respectively on two different logical qubits of toric code. (b) A realistic circuit for $B_{p_0}$ measurement \cite{fowler}. The ancilla qubit is prepared in $\ket{0}$ state, then four $CNOT$ gates are applied in order to couple data and ancilla qubits. Finally, the ancilla is projectively measured in $Z$ basis. (c) A simplified $B_{p_0}$ measurement circuit considered in our work. Note that a five-qubit unitary gate is used here instead of four two-qubit gates. To enable a theoretical analysis of the problem, we focus on the case (c) in our model: (1) Prepare the ancilla qubit in $\ket{+}$ state for each plaquette $p_0$; (2) apply a joint time evolution involving each ancilla and its four neighboring data qubits $\exp [-it Z_{p_0} \otimes B_{p_0}]$ where $Z_{p_0}$ is the Pauli $Z$ acting on ancilla at ${p_0}$; (3) perform projective measurement on ancilla in $Y$ basis. This model is a rather simplified one which is easier to study analytically, but it can capture the fundamental influence of imperfect stabilizer measurement on QEC. 
    %We will show that even such a simplified imperfect measurement model will drastically affect QEC. 
    A more realistic imperfect measurement model relating to (b) will in general has a worse performance, see the discussion of Eq. \eqref{eq:realistic}. A similar construction can also be applied to $A_v$ stabilizers.}
    \label{fig:toric}
\end{figure}

\textbf{Summary of main results:}
In this work, we study the robustness of the toric code QEC~\cite{kitaev}, under the influence of imperfect state preparation and measurements, as visualized in Fig. \ref{fig:toric}, along with stochastic Pauli errors afflicting physical qubits. We consider the realistic case that the measurement apparatus, crucial for both the preparation of the initial state and the subsequent error detection steps, is compromised by coherent noise. The initialization of code states is facilitated through an initial round of stabilizer measurements, following which these states are subjected to an error correction regimen, where the common multi-round syndrome measurements method~\cite{ref:dennis,fowler} is used to decode errors.  Such a procedure is also adopted by experiments \cite{sundaresanDemonstratingMultiroundSubsystem2023}.

Further, we transition the operational framework of this QEC protocol into a novel statistical mechanical (SM) model, characterized by the integration of a 3-dimensional $\mathbb{Z}_2$ gauge theory with a 2-dimensional $\mathbb{Z}_2$ gauge theory. This integration is achieved via a distinctive, non-local timelike coupling originates from the imperfection during initial state preparation. This conceptual model serves as the basis for our exploration into the interaction dynamics between quantum error correction processes and the pervasive influences of coherent noise and stochastic errors, thereby offering novel insights into the resilience and limitations of the toric code QEC methodology under practical conditions. We also emphasize the theoretical novelty in deriving an explicit SM model with coherent errors, as typically, such models are limited to stochastic Pauli errors and Clifford structures.

We find that the Wilson loops in this SM model have an anisotropic behavior: The timelike Wilson loops deconfine at low temperatures (small physical error rates) and confine at high temperatures (large physical error rates); but the spacelike Wilson loops confine at any finite temperature, resulting from the non-local timelike correlation. Then, we predict that there are two thresholds in our QEC model. The confinement-deconfinement transition point of timelike Wilson loops signifies a theoretical threshold located at finite measurement error rate and finite Pauli error rate. However, the confinement behavior of spacelike Wilson loops suggests that the the practical measurement error threshold seats at the point where the initial state preparation is perfect. With arbitrary finite preparation error, the measurement errors will no longer be distinguished from Pauli errors, which results in the failure of QEC even in the limit of large code distance. With a finite code distance $d$, the effectiveness of the pragmatic QEC approach remains viable when the error rate associated with state preparation falls within a region $\sim 1/\log d$. Finally, we emphasize that a more realistic imperfect measurement model relating to Fig. \ref{fig:toric}(b) will in general has a worse performance.

%The structure of this paper is as follows. The error model of imperfect measurement in toric code is presented in Sec. \ref{sec:model} (see Fig. \ref{fig:toric}). In Sec. \ref{sec:SM} we discuss the QEC protocol and its corresponding SM mapping. The statistical phase of the SM model is studied in Sec. \ref{sec:phase}. In Sec. \ref{sec:qec} we decipher the QEC performance according to the phase diagram. In Sec. \ref{sec:3d} we discuss how to increase measurement rounds to prepare logical Pauli bases. In Sec. \ref{sec:realistic} we compare the simplified coherent error model we used with a more realistic error model. Finally, we conclude and have further discussions in Sec. \ref{sec:conclusion}.

%\textbf{{\em Summary of the main results:}} 

\section{Model for toric code under imperfect measurement}
\label{sec:model}

We follow the construction of topological surface code on a torus, i.e. toric code~\cite{kitaev,fowler,ref:dennis,Bombin-arxiv}. 
%It is a stabilizer code defined on a 2D periodic square lattice. 
There are two kinds of stabilizers associated with vertices and plaquettes respectively as shown in Fig.\ref{fig:toric}(a),
\begin{equation}
    A_{v_0} = \prod_{e_0|v_0 \in \partial e_0}X_{e_0},\quad B_{p_0} = \prod_{e_0 \in \partial p_0} Z_{e_0}.
    \label{eq:stabilizers}
\end{equation}
Here we use the symbols $v_0$ and $p_0$ to label vertex and plaquette operators. $X_{e_0}$ and $Z_{e_0}$ represent Pauli operators acting on qubit $e_0$ (i.e. ``edge''). 
%$A_{v_0}$ is the product of four Pauli $X$ operators around vertex $v_0$ and $B_{p_0}$ is the product of four Pauli $Z$ operators around plaquette $p_0$. 
We assume the lattice contains $N$ vertices, $N$ plaquettes and $2N$ edges. $2N$ physical qubits are put on each edge of the lattice. 
Its four-dimensional code subspace is stabilized by all $A_{v_0}$'s and $B_{p_0}$'s which is achieved through projective measurement of these stabilizers. 
Specifically, we start with the logical $++$
 state by projecting all the $B_{p_0}$'s to $+1$ for a product state of physical qubits $\bigotimes_{e_0}\ket{+}_{e_0}$:
\begin{equation}
    \ket{++} = \prod_{p_0} \frac{I+B_{p_0}}{2} \bigotimes_{e_0} \ket{+}_{e_0}
    \label{eq:prepare}
\end{equation}
and other three logical bases are obtained by applying logical Pauli $Z$ operators $\{Z_{l_1}$, $Z_{l_2}\}$ as shown in Fig. \ref{fig:toric}(a),
and these four logical bases form 
a code subspace $\mathcal{C}$.
Correspondingly there are also logical Pauli $X$ operators $\{X_{l_1^*},X_{l_2^*}\}$.
%Here $l_1$ and $l_2$ denote non-contractible loops on the periodic lattice. Logical Pauli $Z$ operators $Z_{l_1}$ and $Z_{l_2}$ are product of $Z$'s along these non-contractible loops. Correspondingly there are also logical Pauli $X$ operators $X_{l_1^*}$ and $X_{l_2^*}$, which are defined as $X$'s along non-contractible loops of the dual lattice, as in Fig. \ref{fig:toric}(a).

Experimentally, the stabilizer measurements are implemented using a multi-qubit unitary operation on a combined qubit set consisting of four data qubits and an ancilla qubit, followed by an ancilla qubit measurement~\cite{fowler,googleScaleUp}, also refer to Fig.\ref{fig:toric}(b).
The correct projective measurements of stabilizers can only be achieved through ideal unitary operations. However, the multi-qubit operation in principle cannot avoid the miscalibration in the experimental setups and results in imperfect measurement~\cite{ref:yang}. Here we consider a simplified imperfect measurement model~\cite{ref:guoyi} with a joint time evolution involving each ancilla and its four neighboring data qubits $\exp [-it Z_{p_0} \otimes B_{p_0}]$  [refer to Fig.\ref{fig:toric}(c)].
%(1) Prepare the ancilla qubit in $\ket{+}$ state for each plaquette $p_0$; (2) apply a joint time evolution involving each ancilla and its four neighboring data qubits $\exp [-it Z_{p_0} \otimes B_{p_0}]$ where $Z_{p_0}$ is the Pauli $Z$ acting on ancilla at ${p_0}$; (3) perform projective measurement on ancilla in $Y$ basis.
Then equivalently we get an operator acting on the data qubits
\begin{equation}
    M_{\{s_{p_0}\}} =  \frac{1}{(\sqrt{2\cosh{\beta}})^N}\exp\left[\frac{1}{2}\beta \sum_{p_0} s_{p_0} B_{p_0}\right].
    \label{eq:imperfect-measurement}
\end{equation}
up to an irrelevant global phase factor. Note that $M_{\{s_{p_0}\}}$ is not a unitary operator.
Here $\tanh(\beta/2)=\tan t$, and $s_{p_0}=\pm 1$ is the measurement outcome of ancilla qubit at $p_0$. We use $\{s_{p_0}\}$ to denote the configuration of ancilla measurement outcomes, which appears with probability $\tr(M_{\{s_{p_0}\}} \rho M_{\{s_{p_0}\}}^\dagger)$ for a given initial state $\rho$ of data qubits.  It is easy to verify that the $E_{\{s_{p_0}\}}= M_{\{s_{p_0}\}}^\dagger M_{\{s_{p_0}\}}$ operators form a set of positive operator-valued measurement (POVM). The error model~\cite{ref:guoyi} only considers the miscalibration of the evolution time $t$ with $0\leq t \leq \pi/4$. When $t=\pi/4$, we have $\beta \rightarrow +\infty$ in Eq. \eqref{eq:imperfect-measurement} and recover the correct projective measurement $M_{\{s_{p_0}\}} \propto \prod_{p_0}(I+s_{p_0}B_{p_0})/2$. 
For $t < \pi/4$ the parameter $\beta$ is finite, and $M_{\{s_{p_0}\}}$ will no longer be a stabilizer projection. 
Here we will assume the coherent deviation $\delta t= t-\pi/4 >0$ to be a fixed value, while in Sec.SVI of supplemental information (SI)~\cite{SupMat}, we briefly discuss the case when it is randomly distributed.
%This situation is referred to as weak measurement in Ref. \cite{ref:guoyi}. 
%Generally, $\beta$ measures how ``strong'' the measurement is, or equivalently how close is our imperfect measurement to the ideal projective measurement. 
Generally, $\beta$
%, termed as ``measurement strength'' in this work,
measures how close is our imperfect measurement to the ideal projective measurement.

%These states are shown to be short-range entangled \cite{ref:guoyi, ref:fisher} by a mapping to the two-dimensional (2D) $\mathbb{Z}_2$ lattice gauge theory, but here we focus on their QEC and error threshold properties.

%In the following parts, we assume the initial code subspace $\mathcal{\widetilde{C}}(\beta)$ is prepared by the aforementioned imperfect measurement.
Experimentally while preparing the initial logical state through stabilizer measurement as in Eq. \eqref{eq:prepare}, the coherent noises are unavoidable, especially for the entanglement gates. For the imperfect measurement model we adopt (Eq. \eqref{eq:imperfect-measurement}), the imperfect initial logical $++$ state is considered as
\begin{equation}
    \ket{\widetilde{++}} = \frac{M_{\{+\}} \bigotimes_{e_0} \ket{+}_{e_0}}{ \sqrt{\bigotimes_{e_0} \bra{+}_{e_0} M_{\{+\}}^\dagger M_{\{+\}} \bigotimes_{e_0} \ket{+}_{e_0}}},
\end{equation}
where all the ancilla measurement outcomes $s_{p_0}$ are set to $+1$. 
We define the other three logical states by applying logical $Z$ operators in analogy to experimental setups, $\ket{\widetilde{-+}}=Z_{l_1}\ket{\widetilde{++}}$, 
$\ket{\widetilde{+-}}=Z_{l_2}\ket{\widetilde{++}}$ and 
$\ket{\widetilde{--}}=Z_{l_1}Z_{l_2}\ket{\widetilde{++}}$. Note that those states are still orthogonal to each other, and define the corresponding code space $\widetilde{\mathcal{C}}(\beta)$ as the subspace spanned by the above four states  (refer to Sec.SI of SI~\cite{SupMat} for more details).
Unlike the perfect measurement case, those logical states depend on the choice of logical operators about where they are located on the physical lattice. However, the key results presented in this work--the statistical mechanical model and its implications to QEC--remain unaffected by the particular selection of logical operators.

These states are argued to be non-topological \cite{ref:guoyi, ref:fisher} by a mapping to the two-dimensional (2D) $\mathbb{Z}_2$ lattice gauge theory. Such states may challenge QEC, yet the behavior of practical QEC procedures and the corresponding error thresholds remains an open question.
Indeed, Ref. \cite{sangMixedstateQuantumPhases2023} suggests that in the presence of stochastic noise, the mixed-state topological phase transition happens before the recoverability transition, given that topological properties are defined by local operations and recovery can involve non-local operations. Additionally, for general imperfect initial codes considered as approximate QEC, Ref. \cite{yiComplexityOrderApproximate2023} identifies non-topological states that correct certain local errors in large systems. Consequently, it is essential to assess a specific QEC protocol with imperfect measurements and directly analyze its error threshold characteristics.

%One might infer that such states will be problematic for QEC, but we still do not know how a practical QEC procedure behaves, and the concrete relation between a state's topological properties and error thresholds remains an open question. 
%In the presence of stochastic noises, Ref. \cite{sangMixedstateQuantumPhases2023} proposed that the mixed-state topological phase transition happens before the recoverability transition, since the topological properties are defined by local operations while recovery could be done by non-local operations. For general imperfect initial codes which could be viewed as approximate QEC, Ref. \cite{yiComplexityOrderApproximate2023} shows there exist states that are non-topological but still correct certain kinds of local errors in the large size limit. Here we focus on a specific QEC protocol with imperfect measurement and directly examine its error threshold properties.

\begin{figure}[t] 
    \includegraphics[width=0.8\columnwidth]{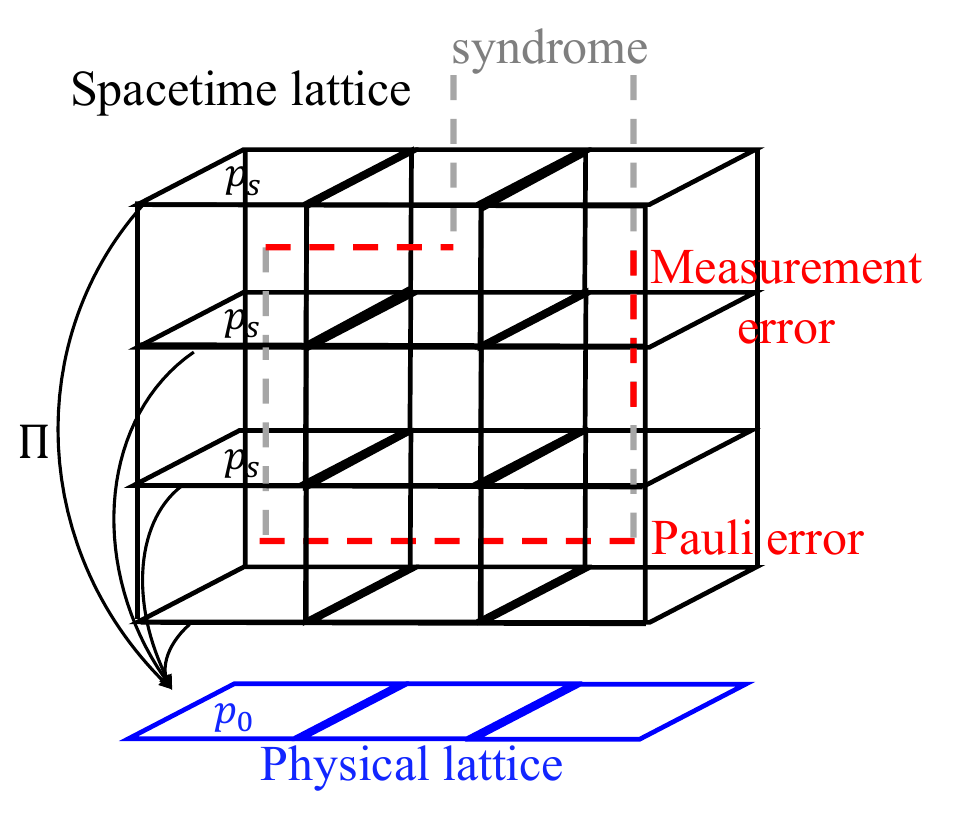}
    \caption{3D spacetime of error history. The black lattice is the spacetime lattice. At the spatial direction, we takes periodic boundary condition; at the temporal direction, we assume that the QEC starts at $t=-\infty$ and ends at $t=+\infty$, such that the 3D corresponding spacetime lattice has infinite boundary condition at time direction. These boundary conditions are also adopted in Ref. \cite{ref:dennis} when discussing SM mapping.
    The Pauli errors, measurement errors and error syndromes are represented as strings on the dual lattice (dashed lines that cross plaquettes). We use timelike and spacelike dual strings to represent measurement errors, i.e. faulty syndromes caused by imperfection of sequential measurements, and Pauli errors, respectively. Given a configuration of Pauli $X$ error (horizontal red strings) for the entire history, the error syndrome (vertical gray strings) will be the configuration of $-1$ ancilla measurement outcomes at different time steps. The error syndrome is supposed to match the endpoints of Pauli error strings, but due to imperfection of syndrome measurements, the syndrome outcomes might be flipped with certain probability. Those flipped syndromes will be referred to as measurement errors (vertical red strings). $\Pi$ denotes the projection from the 3D spacetime lattice to the 2D physical lattice. Given a plaquette $p_s$, $\Pi(p_s)$ yields a plaquette $p_0$ at the same spatial location as $p_s$.}
    \label{fig:spacetime}
\end{figure}

\begin{figure*}
            \centering
            \includegraphics[width=1.85\columnwidth]{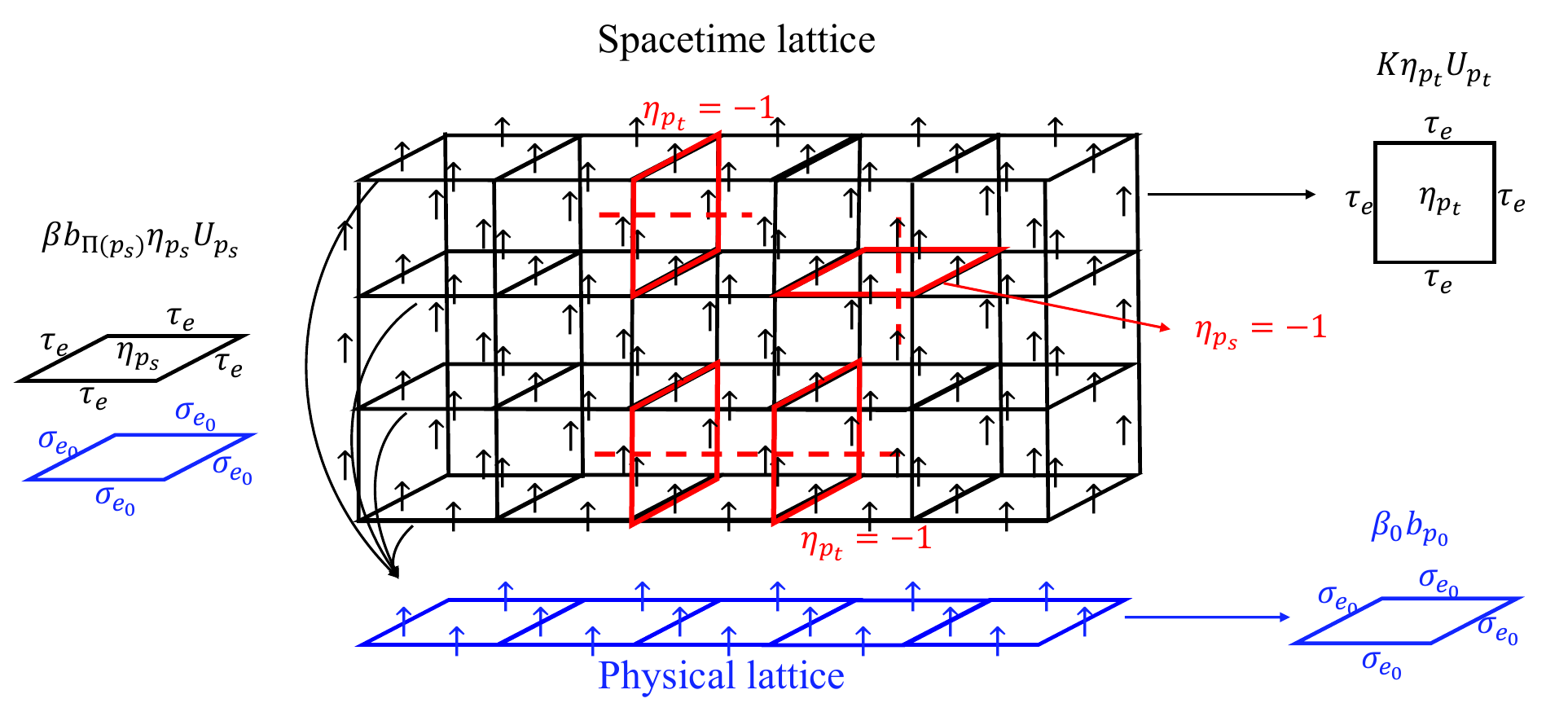}
            \caption{Illustration of the SM model we obtained in Eq. \eqref{eq:partition}. The $\tau_e$ spins are defined on the edges of 3D spacetime lattice and the $\sigma_{e_0}$ spins lie on 2D physical lattice. 
            There are three types of interactions in this model as shown in this figure. $\beta_0 b_{p_0}$ is the gauge interaction term defined on the physical lattice. $K\eta_{p_t} U_{p_t}$ is the timelike gauge interaction on spacetime lattice. The $\beta b_{\Pi(p_s)} \eta_{p_s} U_{p_s}$ term couples the spacelike gauge interaction term $U_{p_s}$ to the gauge interaction term $b_{\Pi(p_s)}$ on physical lattice. The $\eta_p$'s set the signs of gauge interactions on spacetime lattice and they mark the position of error strings during the QEC procedure in Fig. \ref{fig:spacetime}. For example, the flipped interaction $\eta_p=-1$ at plaquette $p$ (red plaquette) corresponds to the presence of an error string at $p$ (dashed red line). The $\{\eta_p\}$ configuration follows a disorder probability Eq. \eqref{eq:prob} that comes from the randomness of Pauli errors and in syndrome measurement. The $\tau_{e}$ spins are non-locally correlated at timelike direction since all spacelike plaquette interactions $U_{p_s}$ along the same timelike arrow are all coupled to the same $b_{\Pi(p_s)}$. Meanwhile, the disorder probability Eq. \eqref{eq:prob} is also correlated at time direction. Physically this is due to the fact imperfection measurement operator will change the current quantum state, which in turn affects subsequent measurement results. It is evident from the expressions Eq. \eqref{eq:partition} and Eq. \eqref{eq:prob} that the non-local correlation results from finite $\beta_0$, or in other words imperfect initial state preparation.}
            \label{fig:gauge-theory}
\end{figure*}

\section{Statistical mechanical mapping}
\label{sec:SM}

Normally for toric code, the Pauli errors are detected by syndrome measurement. However, the syndrome measurement also suffers from imperfection resulting in faulty outcomes. Therefore, in order to distinguish measurement errors from Pauli errors, the standard procedure is to perform multi rounds of syndrome measurements and take into account the obtained entire error history
while decoding~\cite{ref:dennis}. Note that Ref.~\cite{ref:dennis} only consider the stochastic errors of the ancilla measurement and assume a well-prepared initial state from the perfect code space, but we consider imperfect entanglement operations which affect both the initial state preparation and the error detection.
We model the QEC procedure as follows (for convenience we consider only the sector from Pauli $X$ errors): 
\begin{enumerate}
\item[(1)] Start with an arbitrary state $\ket{\widetilde{\Psi}} \in \widetilde{\mathcal{C}}(\beta_0)$, the imperfect measurement strength while preparing the initial state is $\beta_0$.
\item[(2)] Probabilistic Pauli $X$ error acts at each integer valued time $t= -\infty,\cdots,-1,0,1,\cdots,+\infty$. The $X$ error at each physical qubit on each time slice occurs independently with probability $q\in [0,1/2]$. 
\item[(3)] Perform a round of syndrome measurement for each time interval between $t$ and $t+1$. The syndrome measurements are assumed to still suffer from imperfect measurement. So given a configuration of syndrome measurement outcomes $\{s_{p_0}\}$ for a single round, it leads to the action of $M_{\{s_{p_0}\}}$ operator on the current quantum state. Here we set the strength of syndrome measurements to be $\beta$ in order to distinguish them from initial state preparation.
\item[(4)] At the end of the QEC procedure, we decode and apply the Pauli $X$ correction operator to the final state.
\end{enumerate}
We first notice that our model ensures that the imperfection of measurement exclusively affects stabilizer bits while leaving logical information undisturbed, since $M_{\{s_{p_0}\}}$ commutes with logical operators.
So, if the Pauli $X$ errors and the final correction operator compose a contractible loop, 
%(that can be factorized into $B_{p_0} $ stabilizers and causes only trivial effect)
we can verify that the logical information will still be preserved. We refer to this case as the success of QEC, in contrast to the situation with a non-contractible loop causing a logical error. {Note that this condition for successful QEC is the same as the one in Ref. \cite{ref:dennis}.}
%while they only considered the probabilistic ancilla measurement error.

The above QEC procedure could be diagrammatically represented on a three dimensional (3D) cubic lattice in order to decode as in Fig. \ref{fig:spacetime}.
%For convenience, we assume that the QEC starts at $t=-\infty$ and ends at $t=+\infty$, such that the 3D corresponding spacetime lattice has infinite boundary condition at time direction. 
%We use timelike and spacelike dual strings to represent measurement errors, i.e. faulty syndromes caused by imperfection of sequential measurements, and Pauli errors respectively. 
The error strings (including both measurement and Pauli parts) and syndrome strings (marked with $-1$ ancilla outcomes) together compose closed strings or with only endpoints at infinity. The task of the decoder is to identify both measurement and Pauli errors. In order to do so, the decoder should select a configuration of strings (decoding strings) connecting the endpoints of syndrome strings. The timelike (spacelike) parts of the decoding strings represent the measurement (Pauli) error identified by the decoder. QEC succeeds if and only if the decoding strings are topologically equivalent to the real error strings (form contractible loops).
%because that is when the final correction operator and the Pauli $X$ error operators factorize into stabilizers in this setup, as explained in Ref. \cite{ref:dennis}. 
So given an error syndrome, the optimal decoder algorithm~\cite{chubb} should select the topological equivalent class of error strings with the largest probability.

Our main result is that we mapped this QEC scenario to an statistical mechanical (SM) model. 
%Specifically, we derived the probability of a given error configuration.
We denote the vertex, edge and plaquette of the 3D spacetime lattice as $v$, $e$ and $p$. Specifically, the spacelike (timelike) edges and plaquettes will be labeled with a subscript $s$ ($t$), such as $e_s$ and $p_s$ ($e_t$ and $p_t$). We assign a variable $\eta_p=\pm 1$ to each plaquette $p$ to represent the error configuration, e.g. $\eta_p=-1$ for where the measurement or Pauli error presents and $+1$ otherwise. Then the probability of a given error configuration $\{\eta_p\}$ will be 
\begin{align}
    &P(\{\eta_p\})=\nonumber\\
    &\frac{\sum_{\{\sigma_{e_0}\}} \exp\left[ \beta_0 \sum_{p_0} b_{p_0} + \beta \sum_{p_s} b_{\Pi(p_s)}  \eta_{p_s} + K \sum_{p_t} \eta_{p_t} \right]}{4^N (\cosh^N\beta_0 + \sinh^N \beta_0) (2\cosh\beta)^{NT} (2\cosh K)^{2NT}},\label{eq:prob}
\end{align}
where
\begin{equation}
    \quad b_{p_0} = \prod_{e_0 \in \partial p_0} \sigma_{e_0}, \quad K=-\frac{1}{2}\log\frac{q}{1-q}.
\end{equation}
Here $\sigma_{e_0}$ is a classical $\mathbb{Z}_2$ spin-like variable assigned to each edge $e_0$ of the 2D physical lattice. $b_{p_0}$ is the product of four neighbouring $\sigma_{e_0}$'s around the plaquette $p_0$, which has the similar form to a $\mathbb{Z}_2$ gauge interaction term \cite{wegner,ref:kogut}. $\Pi(p_s)$ is the projection of the spacelike plaquette $p_s$ onto the 2D physical lattice, as in Fig. \ref{fig:spacetime}. The label $T$ represents the total number of time steps, and will eventually be taken to $+\infty$. The summation $\sum_{\{\sigma_{e_0}\}}$ runs over all $\{\sigma_{e_0}\}$ configurations.

The detailed derivation of Eq. \eqref{eq:prob} can be found in Sec.SII of SI~\cite{SupMat}, and we only list some key steps here.
Notice that given a quantum state $\rho$, the probability of POVM outcome is $\tr(E_{\{s_{p_0}\}}\rho)$. We may construct a probability of syndrome measurement outcomes of all time steps and space locations conditioned on a fixed Pauli error configuration, which is expressed as 
\begin{equation}
    P\left(\{s_{p_0}(t)\}_t|\{\eta_{e_0}(t)\}_t\right) =|| \prod_t M_{\{s_{p_0}(t)\}} X_{\{\eta_{e_0}(t)\}} \ket{\widetilde{\Psi}} ||^2
    \label{eq:condition-prob1}
\end{equation}
for an arbitrary initial state $\ket{\widetilde{\Psi}}$ in the imperfect code space $\widetilde{\mathcal{C}}(\beta_0)$. Specifically, any initial state should be a superposition of imperfect logical states $\ket{\widetilde{\Psi}} = \Psi_{++} \ket{\widetilde{++}}+\Psi_{+-} \ket{\widetilde{+-}}+\Psi_{-+} \ket{\widetilde{-+}}+\Psi_{--} \ket{\widetilde{--}}$ and we assume it to be normalized. Here we introduced the label $t$ to specify the time step. $\{s_{p_0}(t)\}$ and $\{\eta_{e_0}(t)\}$ denote the syndrome configuration and Pauli error configuration at time $t$, while $\{s_{p_0}(t)\}_t$ and $\{\eta_{e_0}(t)\}_t$ denote the configurations of the whole history. Note that a pair $(p_0,t)$ yields a corresponding spacelike plaquette $p_s$ and a pair $(e_0,t)$ yields a corresponding timelike plaquette $p_t$. $X_{\{\eta_{e_0}(t)\}}$ is the total Pauli error operator at time $t$, and has the form
\begin{equation}
   X_{\{\eta_{e_0}(t)\}} = \prod_{e_0} (\delta_{\eta_{e_0}(t),+1} I +\delta_{\eta_{e_0}(t),-1} X_{e_0}). 
\end{equation}
After combining with Pauli error probability
\begin{equation}
\begin{aligned}
      P(\{\eta_{e_0}(t)\}_t)&=\prod_{e_0,t} q^{\delta_{\eta_{e_0}(t),-1}} (1-q)^{\delta_{\eta_{e_0}(t),+1}}\\
    &=\prod_{e_0,t} \frac{\exp(K\eta_{e_0}(t))}{2\cosh K},
\end{aligned}
\end{equation}
we can obtain the joint probability of total error configurations $P(\{\eta_{p_0}(t)\}_t,\{\eta_{e_0}(t)\}_t)$,
%\begin{equation}
 %   \begin{aligned}
  %  &P(\{\eta_{p_0}(t)\}_t,\{\eta_{e_0}(t)\}_t)= P(\{\eta_{p_0}(t)\}_t|\{\eta_{e_0}(t)\}_t)P(\{\eta_{e_0}(t)\}_t).
    %&= \frac{1}{4^N (\cosh^N\beta_0 + \sinh^N \beta_0) (2\cosh\beta)^{NT}(2\cosh K)^{2NT}}\\ 
    %& \times \sum_{\{\sigma_{e_0}\}} \exp\left[K \sum_{e,t} \eta_{e_0}(t) +  \sum_{p_0} b_{p_0} \left(\beta_0+ \beta\sum_t \eta_{p_0}(t)\right)\right],\\
  %  \end{aligned}
%\end{equation}
which is exactly Eq. \eqref{eq:prob} after converting the notations to those of the 3D lattice (we now denote $\{\eta_p\} = \{\eta_{p_0}(t),\eta_{e_0}(t)\}_t$). 
%Here, the fact $P(\{\eta_{p}\})$ can be derived as a well-defined joint probability for each $\eta_p$ is a consequence of the simpleness of our error model.

Then following the standard procedure described in Ref. \cite{ref:dennis,Bombin-arxiv,chubb}, we computed the probability of the topological equivalent class of error configurations.
by introducing the gauge interaction term $U_p$, the summation over topologically equivalent error configurations is converted to the summation of spin configurations $\{\tau_e\}$ up to a constant factor.
The result is proportional to the partition function of an SM model
\begin{equation}
    \begin{aligned}
        &P([\{\eta_{p}\}]) \propto \mathcal{Z}(\{\eta_p\}) = \sum_{\{\sigma_{e_0}\},\{\tau_e\}} \exp\left[ \beta_0 \sum_{p_0} b_{p_0} \right. \\
        & \left. + \beta \sum_{p_s}  b_{\Pi(p_s)}  \eta_{p_s} U_{p_s} + K \sum_{p_t} \eta_{p_t} U_{p_t} \right],\quad U_{p} = \prod_{e\in \partial p} \tau_e.\\
    \end{aligned}
    \label{eq:partition}
\end{equation}
This SM model is a 3D $\mathbb{Z}_2$ gauge theory defined on the spacetime lattice coupled to a 2D $\mathbb{Z}_2$ gauge theory defined on the physical lattice, see Fig. \ref{fig:gauge-theory}.
Here $\tau_e=\pm 1$ is a spin variable defined on each edge $e$ of the spacetime lattice. $U_p$ is an ordinary $\mathbb{Z}_{2}$ gauge interaction term containing four $\tau_e$ operators, $[\{\eta_{p}\}]$ denotes the topological equivalent class represented by $\{\eta_{p}\}$, and $\eta_p$ sets the sign of interaction term on each plaquette. 
Physically, those $\tau_e$ operators describe the fluctuation of error strings, since flipping $\tau_e$ operator is equivalent to deforming the error strings represented by $\{\eta_{p}\}$. Therefore, the model acquires a local symmetry 
\begin{equation}
    \eta_p \rightarrow \eta_p \prod_{e\in \partial p} \nu_e, \quad \tau_e \rightarrow \tau_e \nu_e, \quad \nu_e = \pm 1,
    \label{eq:gauge-trans}
\end{equation}
which ensures that topologically equivalent error configurations yield the same partition function.

In order to detect the error threshold, Eq. \eqref{eq:prob} can be considered as the quenched disorder probability of interaction configuration $\{\eta_p\}$. Then the phase transition point of $\tau_e$ spins in the disordered SM model corresponds to the error threshold of the QEC model \cite{ref:dennis,Bombin-arxiv,chubb}. 
Note that the 2D gauge model, derived from initial state imperfections \cite{ref:guoyi,ref:fisher}, and the 3D gauge model, characterizing error string fluctuations \cite{ref:dennis}, couple to form a novel system illuminating imperfect initial states' effects on QEC.

%\textcolor{red}{The 2D gauge model originates from the imperfectness of the initial state \cite{ref:guoyi,ref:fisher} and the 3D gauge model describes the fluctuation of error strings \cite{ref:dennis}. Their coupling is a new object that reveals the influence of imperfect initial states on error correction. Since the confinement of the pure 2D gauge model reflects the absence of long-range entanglement \cite{ref:guoyi,ref:fisher}, we expect it to bring non-trivial impacts on QEC while coupled to the 3D gauge model. }

%\textcolor{red}{The 2D gauge model, derived from initial state imperfections \cite{ref:guoyi,ref:fisher}, and the 3D gauge model, characterizing error string fluctuations \cite{ref:dennis}, couple to form a novel system illuminating imperfect initial states' effects on QEC. Given that the confinement in the 2D model signifies the absence of long-range entanglement \cite{ref:guoyi,ref:fisher}, its coupling to the 3D model is anticipated to have non-trivial implications for QEC.}

\begin{figure}
    \includegraphics[width=0.9\columnwidth]{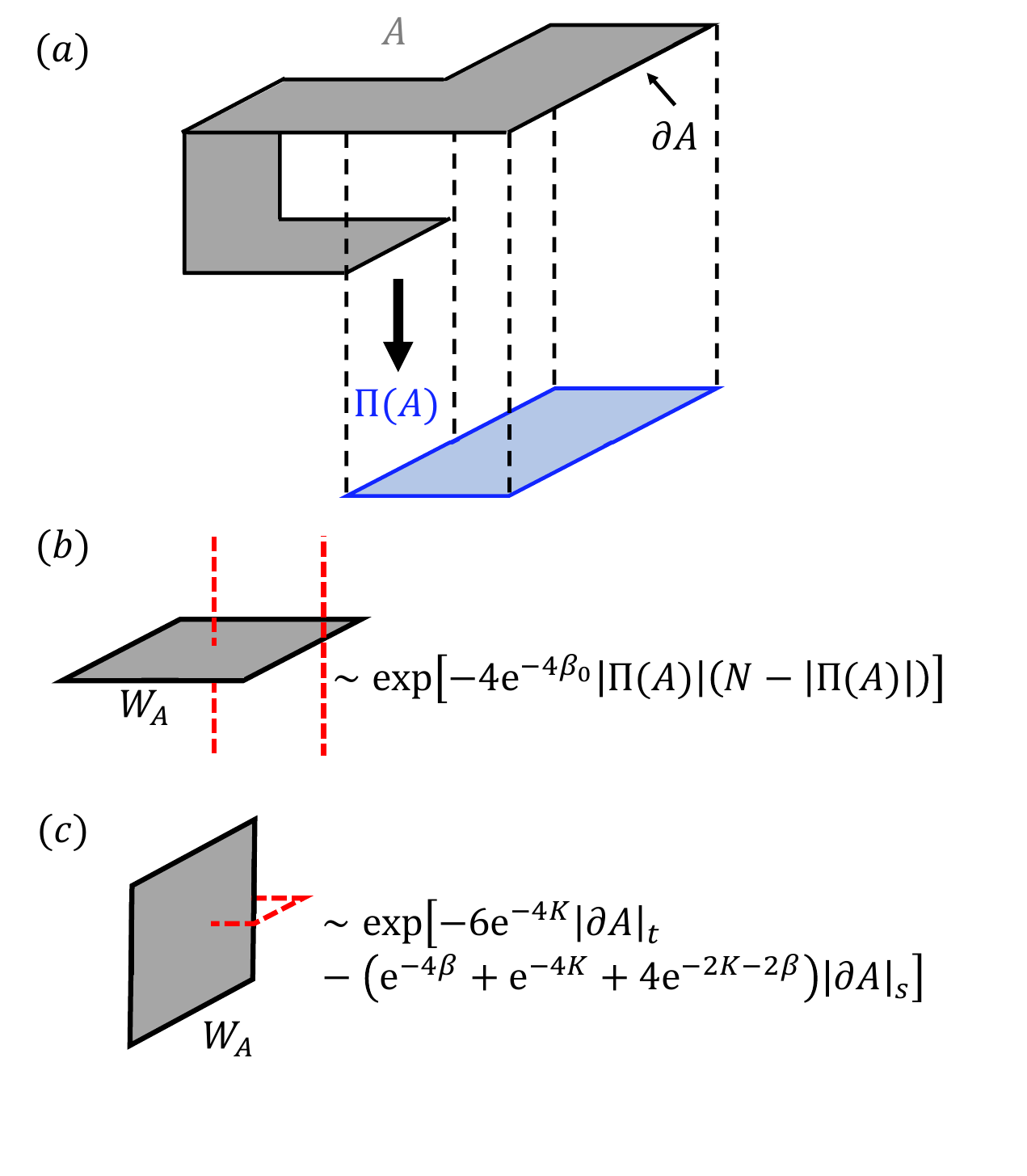}
    \caption{(a) Example of a Wilson loop. $A$ is a surface in 3D spacetime (gray) and $\partial A$ is its boundary (black), which is a closed loop. $\Pi(A)$ is the projection of surface $A$ from 3D spacetime to 2D space mod $\mathbb{Z}_2$. Specifically, under projection the timelike plaquettes are dropped, and an even number of spacelike plaquettes at the same space position also vanishes. The only remaining plaquettes are at the spatial locations that originally have odd numbers of spacelike plaquettes. In (b) and (c), the black lines represent Wilson loops and the red dashed lines are examples of topologically trivial error strings created  by $\tau_e$ fluctuation. (b) A spacelike Wilson loop. In a large enough system it decays exponentially with respect to the area for any finite temperature. 
    %Note that Its scaling behavior written in the figure directly follows from Eq. \eqref{eq:low-T} by constraining $A$ to a pure spacelike region. 
    For a large region $A$, the areal decay will be much faster than the perimetric decay, so the first term in Eq. \eqref{eq:low-T} dominates. This behavior is contributed by the fluctuation of infinite long timelike error strings which are able to appear at any space position (see Sec.SIV of SI~\cite{SupMat}). (c) A timelike Wilson loop. 
    %Its scaling behavior is obtained by constraining $A$ to a pure timelike region in Eq. \eqref{eq:low-T}. 
    Under a low temperature, it decays exponentially with respect to the perimeter. Note that both timelike edges and spacelike edges are contained in the boundary of a timelike region, determining the $|\partial A|_t$ term and $|\partial A|_s$ term respectively. 
    %This perimetric decay behavior is mainly contributed by local error loops near the Wilson loop (see Sec.SIV of SI~\cite{SupMat}).
    }
    \label{fig:wilson}
\end{figure}
%Here we provide some analytical results about the SM model and its phase structure. 

\section{Phase structure of the statistical mechanical model}
\label{sec:phase}
First, we notice that the SM model has a non-local correlation at time direction originates from imperfect initial state preparation, see Fig. \ref{fig:gauge-theory}.
If the initial state is well prepared ($\beta_0\rightarrow\infty$), the code space $\widetilde{\mathcal{C}}(\beta_0)$ becomes exactly the toric code subspace. The action of following syndrome measurement operators Eq. \eqref{eq:imperfect-measurement} on this space yields only a global phase factor and does not change the state itself. 
%In this case, even though the measurement outcome can still be faulty, the probability of measurement error will now become uncorrelated. 
In this case, the model reduces to the familiar random plaquette gauge model (RPGM)~\cite{ref:dennis,ref:preskill-numerical,ref:numerical}. 
%The phase structure of RPGM is discussed in Ref. \cite{ref:dennis,ref:preskill-numerical}.

%Specifically, when $\beta_0 \rightarrow +\infty$, all the $b_{p_0}$'s will be set to $+1$ in Eq. \eqref{eq:partition} and we arrive at:
%\begin{equation}
%    \begin{aligned}
%    &\mathcal{Z}(\{\eta_p\}) = \sum_{\{\sigma_{e_0}\},\{\tau_e\}} \exp\left[\beta \sum_{p_s} \eta_{p_s} U_{p_s} + K \sum_{p_t} \eta_{p_t} U_{p_t} \right],\\
%    &U_{p} = \prod_{e\in \partial p} \tau_e,\\
%    &P(\{\eta_e\})= \frac{\exp\left[K \sum_{p_t} \eta_{p_t} + \beta \sum_{p_s} \eta_{p_s} \right]}{\prod_t [(2\cosh\beta)^{N} (2\cosh K)^{2N}]} ,\\ 
%    \end{aligned}
%    \label{eq:preskill-model}
%\end{equation}
%which has the same structure as the one in Ref. \cite{ref:dennis} to describe the error threshold under probabilistic measurement and Pauli errors. \\

In reality, the faulty circuits that produce the imperfect syndrome measurements also provide the imperfect initial state preparations.
In this situation, i.e. with finite $\beta_0$, the non-local timelike correlation will lead to a different phase structure in stark contract to RPGM. In order to detect the phase diagram of this model, we consider the Wilson loop
\begin{equation}
    W_A = \prod_{p \in A} U_p = \prod_{e\in \partial A} \tau_e,
\end{equation}
which serves as the order parameter for $\mathbb{Z}_2$ gauge theory. 
Here $A$ is a set of plaquettes representing a surface in spacetime. The product of $U_p$'s on surface $A$ equals the product of $\tau_e$'s on $\partial A$, which is the boundary of surface $A$ and forms a closed loop. 
In the conventional $\mathbb{Z}_2$ gauge theory \cite{wegner,ref:kogut} the scaling behavior of Wilson loop expectation values with respect to the loop size distinguishes between the confinement (disordered) phase and the deconfinement (ordered) phase.
In the deconfinement phase, it decays exponentially with respect to the perimeter of the loop, 
\begin{equation}
    W_A \sim \exp(-const\times |\partial A|),
    \label{eq:perimeter-law}
\end{equation}
called perimeter law. Here we use $|\cdot|$ to denote the cardinal of a set (i.e. the number of the elements of a set). For example $|\partial A|$ is the number of edges contained in $\partial A$. On the other hand, in the confinement phase, the scaling behavior of Wilson loops obeys area law, 
\begin{equation}
    W_A \sim \exp(-const\times |A_{min}|),
    \label{eq:area-law}
\end{equation}
where $A_{min}$ is the minimal surface enclosed by $\partial A$.
Here we will study the expectation value of Wilson loops in our SM model. %relate wilson loop to phases

Note that our SM model satisfies a generalized version of Nishimori condition \cite{ref:nishimori}, which means that the error rate parameters $(\beta_0, \beta, K)$ in the quenched disorder probability in Eq. \eqref{eq:prob} are the same as those in the partition function in Eq. \eqref{eq:partition}, respectively.
%If we set the parameters in Eq. \eqref{eq:prob} to be $(\beta_0', beta', K')$ that is different from $(\beta_0, beta, K)$ in Eq. \eqref{eq:partition}, then the disordered SM model deviates from Nishimori condition and does not correspond to the QEC procedure anymore.
Under this condition, by taking advantage of a local symmetry of the model shown in Eq. \eqref{eq:gauge-trans}, we find that (refer to Sec.III of SI~\cite{SupMat})
\begin{equation}
    [\braket{W_{A}}] = [\braket{W_A}^2].
\end{equation}
The above equality suggests the absence of gauge glass phase \cite{ref:preskill-numerical}, where $[\braket{W_{A}}]$ obeys area law Eq. \eqref{eq:area-law} but $[\braket{W_{A}}^2]$ obeys perimeter law Eq. \eqref{eq:perimeter-law}. In that case, we only need to concern about the deconfinement-confinement phase transition of $\tau_e$'s under Nishimori condition. 

We then perform a low-temperature (i.e. $\beta_0$, $\beta$ and $K$ are sufficiently large, corresponding to small enough physical error rates) expansion \cite{ref:kogut} for $[\braket{W_A}]$. Here, $\braket{\cdot}$ denotes the ensemble average with respect to the model Eq.\eqref{eq:partition} under a specific configuration; and $[\cdot] = \sum_{\{\eta_p\}} P(\{\eta_p\}) (\cdot)$ represents the disorder average over different configurations with respect to the probability Eq.\eqref{eq:prob}. 
%Here low temperature means that the parameters $\beta_0$, $\beta$ and $K$ are sufficiently large, corresponding to small enough physical error rates. 
We assume $\e^{-\beta_0}$, $\e^{-\beta}$ and $\e^{-K}$ are of the same order and expand $\log [\braket{W_A}]$ up to the first non-vanishing order $\e^{-4\beta_0}$. We obtain the result
\begin{equation}
\begin{aligned}
&[\braket{W_A}] \simeq \exp[ - 4 \e^{-4\beta_0}|\Pi(A)|(N-|\Pi(A)|)\\
&-  \left(\e^{-4\beta} + \e^{-4K} + 4\e^{-2\beta - 2K} \right)|\partial A|_s - 6  \e^{-4K} |\partial A|_t].\\
\end{aligned}
\label{eq:low-T}
\end{equation}
Here $\Pi$ is defined as projection from 3D spacetime to 2D space mod $\mathbb{Z}_2$, illustrated in Fig. \ref{fig:wilson}(a). $|\partial A|_s$ ($|\partial A|_t$) denotes the spacelike (timelike) edges $e_s$ ($e_t$) contained in $\partial A$. 
The derivation of the expansion is lengthy and refer to Sec.SIV of SI~\cite{SupMat} for details.
%Breifly it is done by first expanding $P(\{\eta_p\})$ up to $\e^{-4\beta_0}$. Then for each error configuration that appeared in the expansion, we compute the expectation value $\braket{W_A}$ up to the order we need. The perturbative evaluation of $\braket{W_A}$ is accomplished by identifying the ground state and then taking the lowest excited states into consideration. Each of these states yields a specific value of $W_A$. Putting all these things together, we obtain Eq. \eqref{eq:low-T}. 

%From this expression, we see that the expectation value of Wilson loops has an anisotropic scaling behavior. It deconfines at the timelike direction under low temperature but confines at the spacelike direction for any finite $\beta_0$ (Fig. \ref{fig:wilson}). At a sufficiently high temperature (or large error rate), the system should be drived into a completely disordered phase, so we expect a phase transition to a regime that confines the timelike Wilson loop. We also noticed from Eq. \eqref{eq:low-T} that the  areal decay at low temperatures is a consequence of imperfect measurement during initial state preparation. In comparison, if the initial state is ideally prepared ($\beta_0=+\infty$, corresponds to RPGM), the Wilson loops will acquire an isotropic scaling behavior, i.e. both the spacelike and timelike Wilson loops will exhibit perimetric decay at the low-temperature phase and areal decay at the high-temperature phase \cite{ref:dennis}. The qualitative phase diagram is summarised in Fig. \ref{fig:finite} (a).

\begin{figure*}
    \centering
    \includegraphics[width=2\columnwidth]{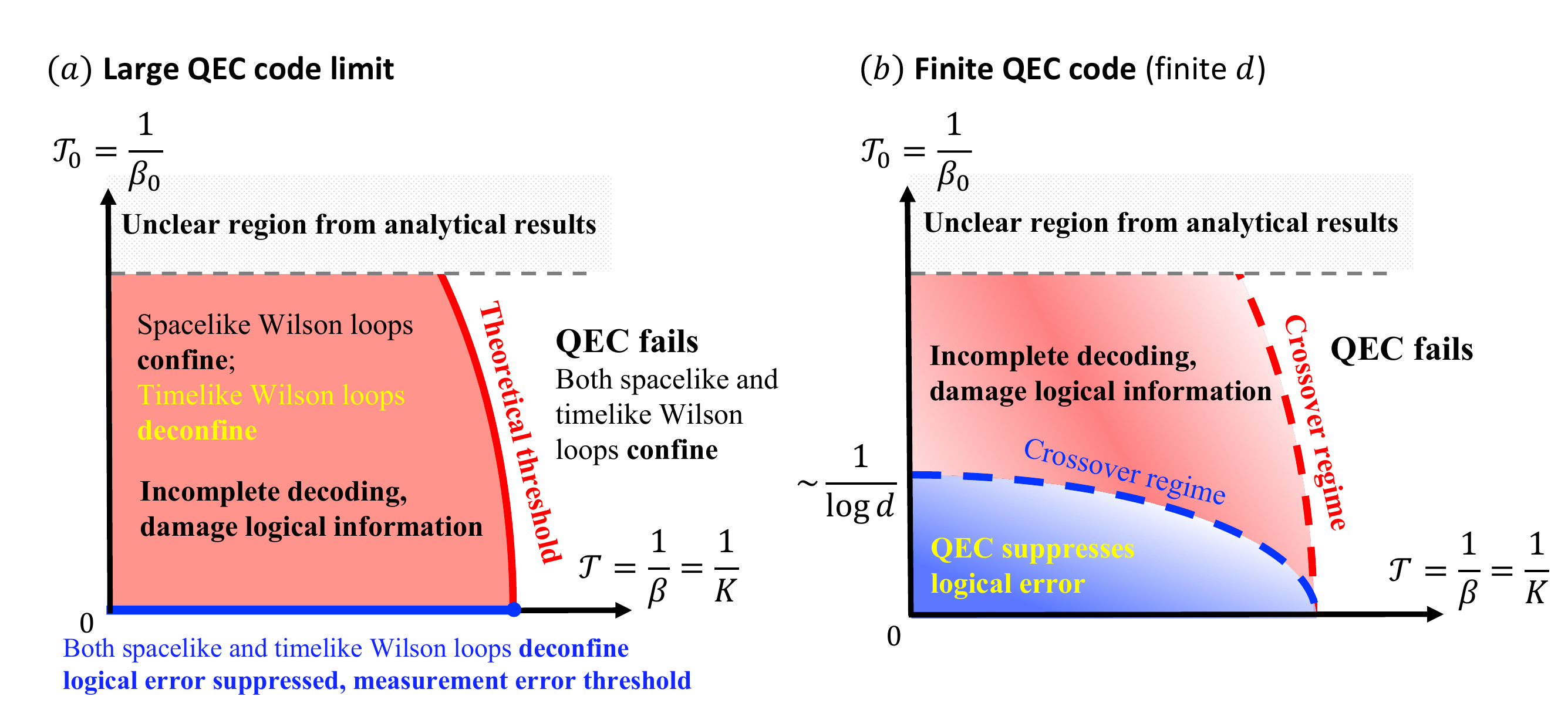}
    \caption{(a) Our estimation of the phase structure in the thermodynamic limit $T\rightarrow +\infty$ and $N\rightarrow +\infty$ while setting $\mathcal{T} = 1/\beta = 1/K$ and $\mathcal{T}_0 = 1/\beta_0$. Above the theoretical threshold (red line) QEC fails due to non-contractible logical Pauli errors. 
    Below the theoretical threshold and above the measurement error threshold (blue line), non-contractible logical Pauli errors are suppressed. However, measurement errors are still unidentifiable through decoding a finite error history and will be confounded with Pauli errors. Note that the $\mathcal{T}$ axis where $\mathcal{T}_0=0$ represents the RPGM. While the theoretical threshold intersect with the $\mathcal{T}$ axis at the common RPGM phase transition point \cite{ref:dennis}, we are not sure yet whether it intersect with the $\mathcal{T}_0$ axis. Some details at higher temperatures still requires further investigation.  (b) The phase diagram while fixing a finite code distance $d$. The phase transitions (thresholds) in Fig. \ref{fig:finite}(a) are smoothed into crossovers due to finite-size effect. Especially, the measurement error threshold becomes a finite-temperature crossover (blue dashed line), leading to a parameter region with a finite area (light blue region) that effectively suppresses logical errors. The parameters in this region should satisfy either Eq. \eqref{eq:distance1} or Eq. \eqref{eq:distance2} such that the non-local measurement errors will not be a problem. Specifically, the crossover condition $\mathcal{T}_0 \sim 1/\log d$ near the $\mathcal{T}_0$ axis is derived from Eq. \eqref{eq:distance1} (However, Eq. \eqref{eq:distance1} or Eq. \eqref{eq:distance2} are only approximate expressions valid in the low-temperature limit. The precise value of this crossover still requires further investigation.). Increasing $d$, this region becomes smaller and smaller and eventually sticks to the $\mathcal{T}$ axis. Above the crossover regime of measurement threshold in the light red region, the effect of non-local measurement errors on the QEC becomes non-negligible. As for the SM model side, the blue crossover detects the confinement of spacelike Wilson loops while the red crossover detects the confinement of timelike Wilson loops.}
    \label{fig:finite} 
\end{figure*}

Eq. \eqref{eq:low-T} provides us with insights into the phase structure. As we have mentioned, the Wilson loops have anisotropic scaling behavior at low temperatures. A pure timelike Wilson loop $W_A$ which contains only timelike plaquettes is shown in Fig. \ref{fig:wilson}(c). It deconfines and decays exponentially with respect to perimeter as in a conventional 3D $\mathbb{Z}_2$ lattice gauge theory under low temperature. Meanwhile, a pure spacelike Wilson loop is shown in Fig. \ref{fig:wilson}(b). For large enough system size $N$, its areal decay is faster than the perimetric decay, so the first term in Eq. \eqref{eq:low-T} dominates and $[\braket{W_A}]$ confines as long as the temperature is finite. We notice that no matter how low the non-zero temperature is, the confinement is always maintained. In addition, a sufficiently high temperature should always drive the system into a completely disordered phase, which will confine all Wilson loops. Thus the spacelike Wilson loops confine at any finite temperature (or error rate). In comparison, if the initial state is ideally prepared ($\beta_0=+\infty$, corresponds to RPGM), the Wilson loops will acquire an isotropic scaling behavior, i.e. both the spacelike and timelike Wilson loops will exhibit perimetric decay at the low-temperature phase and areal decay at the high-temperature phase \cite{ref:dennis}. The qualitative phase diagram is summarised in Fig. \ref{fig:finite} (a).

%Here we also remark that the low-temperature expansion result shown in Eq. \eqref{eq:low-T} is valid for any finite system size $N$ and region $A$, as long as the temperature (physical error rate) is sufficiently low. The subtlety is that we cannot directly take the thermodynamic limit $N\rightarrow +\infty$ in Eq. \eqref{eq:low-T} because of the factor $N$ contained in the leading order. Actually, the appearance of area term $|\Pi(A)|(N-|\Pi(A)|)$ in Eq. \eqref{eq:low-T} is a natural result because our space manifold is a closed surface (due to the periodic boundary condition), thus $\Pi(A)$ and its complement on 2D space yield the same boundary, and should be symmetric in an expression containing $|\Pi(A)|$. However under the thermodynamic limit, the phase structure of the SM model should not depend on this boundary condition, and we expect that the area law is still obeyed by spacelike Wilson loops in the low-temperature phase. A brief discussion about the analogy to the exactly solvable 2D $\mathbb{Z}_2$ lattice gauge theory can be found in Sec. SIV of SI~\cite{SupMat}.

\section{Impact on quantum error correction}
\label{sec:qec}
%Previously, we mapped our QEC model with imperfect measurement and Pauli error to an SM model and analyzed its phase structure. 
%We now discuss their influences on QEC performance and threshold theorem. 
Following the mapping of our QEC model to an SM model, we now examine its implications on QEC performance and the threshold theorem.

\begin{table*}
    \centering
    \includegraphics[width=2\columnwidth]{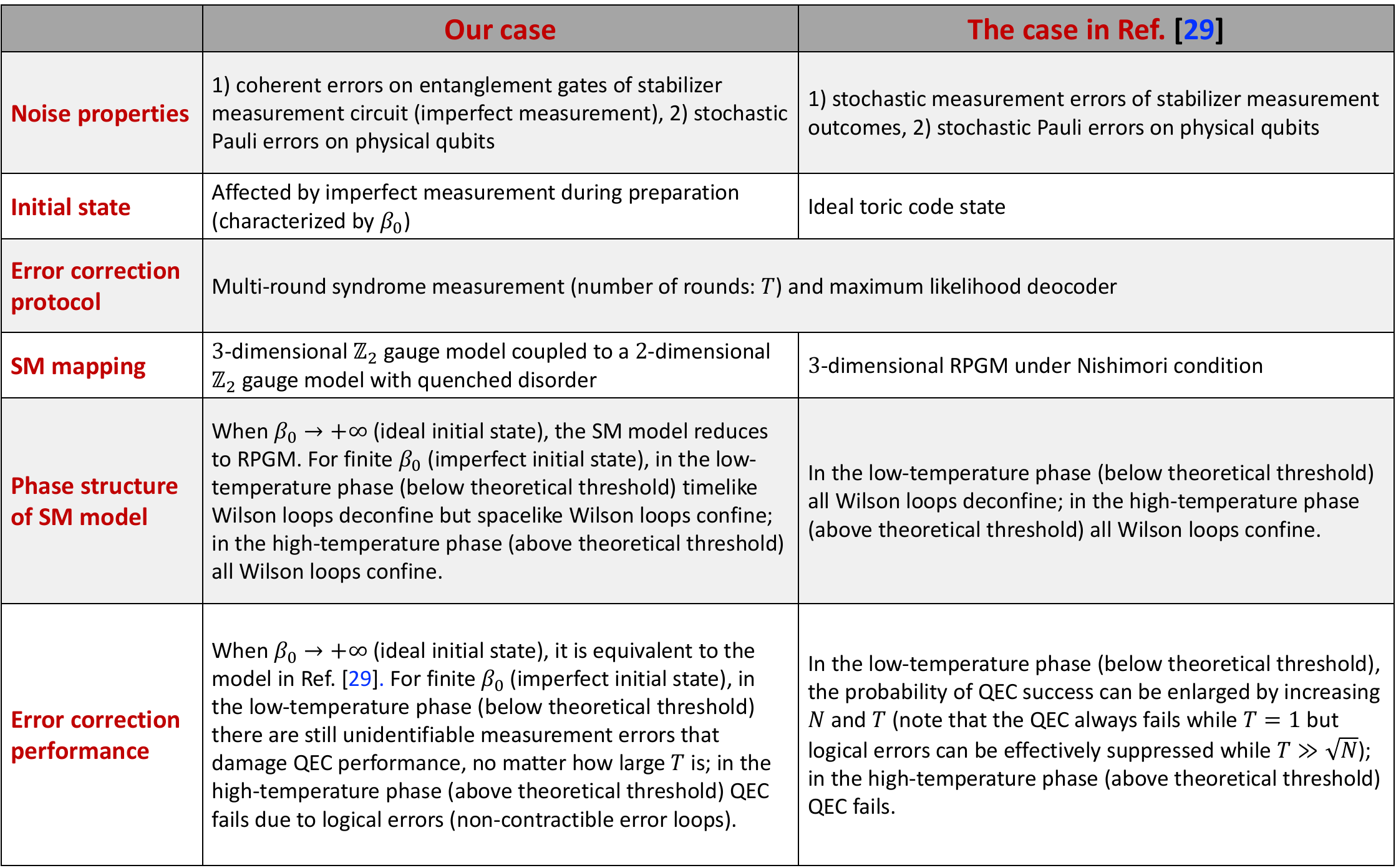}
    \caption{Comparison between our case and the case with the stochastic measurement error model \cite{ref:dennis}. Note that although the measurement noises of the two models are different at the physical level, their error correction properties and corresponding SM models will become equivalent when we set the initial state in our model to be ideal ($\beta_0 \rightarrow +\infty$).}
    \label{tab:compare} 
\end{table*}
\subsection{Behavior in an infinite size system }
First of all, recognizing the existence of a confinement transition point concerning timelike Wilson loops in the limit $T\rightarrow +\infty$ and $N\rightarrow +\infty$, 
we ascertain that it also separates different behaviors of logical error rate. In fact, we find that the logical errors associated with spacelike non-contractible loops are suppressed at the low-temperature phase, indicating a transition we refer to as the theoretical threshold. 
In our derivation of low-temperature expansion (Sec.SIV of SI~\cite{SupMat}), we find that: 
\begin{enumerate}
\item The decay in area law for spacelike Wilson loops is attributed to the emergence of non-local timelike error strings as depicted in Fig. \ref{fig:wilson}(b); 
\item At sufficiently low temperatures, non-local timelike strings and local error loops manifest with relative independence, resulting in the confinement of local error loops, thereby preventing their extension to arbitrary lengths.
\end{enumerate}
%the non-local timelike strings and other local error loops appear in a relatively independent manner at sufficiently low temperatures. Therefore, those local error loops are compressed and are unlikely to stretch to arbitrarily large. 
Specifically, this implies that fluctuating error strings are constrained (or unable to extend indefinitely) in their spatial extension, leading to a negligible error rate for spacelike non-contractible loops or logical Pauli operators. 
By examining the domain wall free energy cost \cite{ref:dennis} associated with these non-contractible loops, it is found to be proportionate to the code distance $d$ within the red phase as shown in Fig. \ref{fig:finite}(a).
%If we measure the domain wall free energy cost \cite{ref:dennis} of those non-contractible loops, we will find that it is still proportional to the code distance $d$ in the red phase Fig. \ref{fig:finite}(a). 
By increasing the temperature,  a transition point is anticipated where timelike Wilson loops become confined, concurrently with the proliferation of spacelike non-contractible error strings.
%there should be a transition point where timelike Wilson loops confine and simultaneously spacelike non-contractible error strings proliferates. 
However, this threshold, which will be elucidated subsequently, does not encompass the correctability of measurement errors. 
Above this theoretical threshold, QEC is deemed ineffective, as evidenced by a persistent and finite logical error rate that is not ameliorated by enlarging the system size, particularly within the red phase depicted in Fig. \ref{fig:finite}(a).
Recall that confinement of spacelike Wilson loops is observed even below the theoretical threshold due to the fluctuation of timelike non-local error strings (Fig. \ref{fig:wilson}(b)). Although those timelike error strings do not directly relate to logical Pauli operators, they signify the proliferation of measurement errors. 
Those non-local measurement errors (extending the whole time interval $[0,T]$ for any finite $T$) become indistinguishable from local Pauli errors and thus damage the decoding process.
For example, the true syndrome of a single Pauli error string will be two non-local timelike strings starting from its boundary points. But in the meantime, the syndrome can be generated by non-local measurement errors, which may occur with a probability comparable to that of the Pauli string and are impervious to suppression by large $T$ or large $d$. The consequence is that the decoder might mix up these two situations with a finite probability and inappropriately omit the correction of the aforementioned Pauli error. If this Pauli error anti-commutes with logical $Z$ operators, the logical information will be damaged. Note that in this case, the defects caused by faulty measurement and Pauli error complements each other in the physical states, so the error cannot be detected even when applying more error correction cycles. 
In the low-temperature limit, we expect the logical error rate (measured with worst-case infidelity) to be proportional to the code distance, see Sec.SV of SI~\cite{SupMat}. 
Given that the non-local timelike error strings consistently perforate the entire lattice, the choice of $T$ in reality for decoding is inconsequential.
Hence, the logical error rate is also not suppressed by increasing $T$. These results indicate the loss of quantum information.

%\textcolor{red}{Consequently, one key outcome of our investigations can be summarized as follows: a synergistic interaction between timelike non-local measurement errors and local Pauli errors introduces a novel mechanism for the induction of logical errors. Crucially, the probability of such error occurrences does not diminish with increases in either the temporal duration, $T$, or the system size, $d$. In instances where $\beta_0$ remains finite, the QEC framework exhibits characteristics akin to those observed when surpassing the genuine error threshold. Note that this mechanism is not inherently apparent within an idealized infinite-time decoding framework. Within such theoretical constructs, non-local timelike syndromes are hypothesized to be interconnected by error strings extending to both $t=+\infty$ and $t=-\infty$, thus forming uninterrupted world lines. This theoretical model, however, diverges from practical decoding processes, which are limited to considering a finite history of errors. This divergence highlights a critical limitation in the existing conceptualization of QEC strategies, necessitating a reevaluation to incorporate the practical constraints of temporal finitude.}

Henceforth, our investigation elucidates crucial findings: Timelike non-local measurement errors and local Pauli errors together constitute a new mechanism that causes logical errors, and the probability of those events cannot be suppressed by large $T$ or $d$. So when $\beta_0$ is finite, the QEC system behaves like it is above the true error threshold. Notice that the above mechanism is not explicit in the infinite time decoding scenario, since the non-local timelike syndromes are imagined to be connected by error strings at both the $t=+\infty$ and $t=-\infty$ ends, constituting complete world lines. This is not true in a realistic decoding process, because only finite error history can be taken into consideration.

This is in stark contrast to the case with only stochastic errors~\cite{ref:dennis}, where the perimeter law for both the spacelike and timelike Wilson loops guarantees the achievement of an effective error correction with a finite error history. 
Furthermore, it is worth noting that the RPGM \cite{ref:dennis} manifests an isotropic characteristic, i.e. the simultaneous deconfinement and confinement of the spacelike and timelike Wilson loops respectively, hence, obviating the need for distinct analyses of various error types.
In our case, the inability to correct measurement errors is caused by the finite value of $\beta_0$, i.e. the imperfection of initial state preparation; so we refer to $\beta_0 \rightarrow +\infty$ as the measurement error threshold of our error model. 
Setting $\beta = K$, a sketch of the phase diagram is shown in Fig. \ref{fig:finite}(a). In addition, a comparison between our error model and that with only stochastic errors~\cite{ref:dennis} is shown in Tab. \ref{tab:compare}.

For completeness, we scrutinize the phase structure in various limits.
%$K \rightarrow +\infty$ or $\beta \rightarrow +\infty$.
Notice that the red phase in Fig. \ref{fig:finite}(a) signifies the proliferation of measurement error strings, but logical errors result from a combination of measurement errors and Pauli errors.
In the limit $K\rightarrow +\infty$ (keeping $\beta_0$ and $\beta$ finite), the QEC protocol trivially succeeds as there will be no Pauli errors. But as long as $K$ is finite, the logical error rate is not suppressed by the system size due to the indistinguishability between measurement errors and Pauli errors. In the scenario where $\beta \rightarrow +\infty$ but $K$ and $\beta_0$ are finite, the non-local measurement errors are still present and cannot be decoded when intertwined with Pauli errors. As for the $\beta \rightarrow_0 +\infty$ limit, it reduces to the RPGM as mentioned before.

\subsection{Finite-size effect}
For a small code with limited system size $N$, we infer from Eq. \eqref{eq:low-T} that the above problem might be circumvented with the limit
\begin{equation}
    d \ll \e^{\beta_0},
    \label{eq:distance1}
\end{equation}
or 
\begin{equation}
    d \ll [\e^{4\beta_0} (\e^{-4\beta}+\e^{-4K} + 4\e^{-2\beta-2K})]^{1/3}.
    \label{eq:distance2}
\end{equation}
Here $d$ is the code distance and $N=d^2$.
The first bound is derived by assuming the areal term in Eq. \eqref{eq:low-T} is negligible
\begin{equation}
    e^{-4\beta_0} |\Pi(A)|(d^2-|\Pi(A)|) \ll 1,
\end{equation}
for all spacelike Wilson loops $A$. The left-hand-side maximizes when $A$ is half the size of the spatial lattice $|\Pi(A)| = d^2/2$. Substituting $|\Pi(A)| = d^2/2$ into the above expression, we have
\begin{equation}
    e^{-4\beta_0} d^4/4 \ll 1,
\end{equation}
which is equivalent to Eq. \eqref{eq:distance1} ignoring a constant factor.
Physically, Eq. \eqref{eq:distance1} is interpreted as the probability of the appearance of non-local measurement errors on the whole system is ignorable.
The second bound is derived by considering the case when the perimetric decay is faster than the areal decay in Eq. \eqref{eq:low-T} for a spacelike Wilson loop,
\begin{equation}
\begin{aligned}
    &    e^{-4\beta_0} |\Pi(A)|(d^2-|\Pi(A)|)\\
    &\ll \left(\e^{-4\beta} + \e^{-4K} + 4\e^{-2\beta - 2K} \right)|\partial A|_s.
\end{aligned}
\end{equation}
Still we require $A$ to be half of the spatial lattice, $|\Pi(A)| \sim d^2$ and $|\partial A|_s \sim d$. Thus we have 
\begin{equation}
    e^{-4\beta_0} d^4 \ll \left(\e^{-4\beta} + \e^{-4K} + 4\e^{-2\beta - 2K} \right)d,
\end{equation}
which leads to Eq. \eqref{eq:distance2}.
Physically, Eq. \eqref{eq:distance2} is interpreted as the influence of non-local error strings is not significant compared to other local error strings.
If either of these two bounds is satisfied, we anticipate that the ability of our QEC procedure to detect measurement errors will be similar to that of Ref. \cite{ref:dennis}. To do so, the imperfection of initial state preparation must be negligible or much smaller than the syndrome measurement imperfection and Pauli error rate. Even then the code distance is still upper bounded if we fix the error parameters $\beta_0$, $\beta$ and $K$. Usually, when performing QEC, we anticipate increasing code distance to suppress the logical error rate \cite{fowler}. However, for the important error problem considered here, Eq. \eqref{eq:distance1} and Eq. \eqref{eq:distance2} form bounds that prevent the code from scaling up. Equivalently if we fix $d$ and vary the error parameters, we obtain the phase diagram in Fig. \ref{fig:finite}(b). It is noteworthy that the region enabling pragmatic error correction only experiences a gradual reduction as increasing the code distance, i.e. $\sim 1/\log d$, which is not excessively frustrating. In Sec.SVII of SI~\cite{SupMat}, we describe a preparation procedure with the multiple-round measurement
protocols that maintains decodability in small finite code systems.

%In this work, we discuss the imperfect measurement problem based on the circuit shown in Fig. \ref{fig:toric}(c) (e.g. for $B_{p_0}$), which allows us to conduct an analytic study. 
%By mapping the standard QEC procedure under imperfect measurement to an SM model, we find two phases that have different QEC performances. The high temperature (high physical error rate) phase signifies the failure of QEC caused by non-contractible error strings. In the low temperature (low physical error rate) phase, the measurement errors cannot be identified through decoding syndrome outcomes of finite rounds. 

\section{Relation to a realistic measurement circuit}
 \label{sec:realistic}
In fact, the circuit shown in Fig. \ref{fig:toric}(b), which contains only two-qubit gates rather than a five-qubit evolution in our simple model, is more realistic, and it is expected to have worse performance while suffering from coherent noise on entanglement gates. 
Ref. \cite{ref:yang} discussed an imperfect measurement model that mimics the behavior of superconducting quantum computation systems. The $CNOT$ gate is divided into a $CZ$ gate and two Hadamard gates, $CNOT = H( CZ )H$ where $H$ is the Hadamard gate acting on the target qubit (ancilla qubit in our setup). Each $CZ$ gate is implemented by a time evolution
\begin{equation}
    U = \exp\left[-\ii \frac{t}{4} \left(s^{z}\otimes \sigma_i^z - s^z \otimes I - I \otimes \sigma^z_i + I \otimes I\right)\right]
\end{equation}
Here $s$ labels the ancilla qubit and $\sigma_i$, $i=1,2,3,4$ labels the four data qubits. It recovers the $CZ$ gate when $t=\pi$. Assume the final ancilla measurement has an outcome $s=\pm 1$, the corresponding action on data qubits is
\begin{widetext}
\begin{equation}
    \begin{aligned}
        &M_s = \bra{s} H \exp\left[-\ii \frac{t}{4} \sum_i \left(s^{z}\otimes \sigma_i^z - s^z - \sigma^z_i + I \right)\right] H \ket{0}
        %&= \frac{1}{2} \left(1+s \e^{-\ii 2t} \cos^4\frac{t}{2}   +\ii s \e^{-\ii 2t} \sin\frac{t}{2}\cos^3\frac{t}{2} \sum_{i} \sigma^z_i \right.\\
        %& \left.- s \e^{-\ii 2t} \sin^2\frac{t}{2}\cos^2\frac{t}{2} \sum_{i<j} \sigma^z_i \sigma^z_j -\ii s \e^{-\ii 2t} \sin^3\frac{t}{2}\cos\frac{t}{2} \sum_{i<j<k} \sigma^z_i \sigma^z_j \sigma^z_k + s \e^{-\ii 2t} \sin^4\frac{t}{2} B_{p_0}\right)\\
        =\frac{1}{2}\left( 1 +s \e^{-\ii 2t} \cos^4\frac{t}{2} + s \e^{-\ii 2t} \sin^4\frac{t}{2} B_{p_0}\right)\\
        &\times\left(1 - \frac{\ii s \e^{\ii 2t} \sin\frac{t}{2}\cos^3\frac{t}{2} + \ii \sin\frac{t}{2}\cos^7\frac{t}{2} + \ii \sin^7\frac{t}{2}\cos\frac{t}{2}}{\left(s \e^{\ii 2t} + \cos^4\frac{t}{2}\right)^2 -  \sin^8\frac{t}{2} } \sum_i \sigma^z_i - \frac{\sin^2\frac{t}{2} \cos^2\frac{t}{2}}{s \e^{\ii 2t} + \cos^4\frac{t}{2} +\sin^4\frac{t}{2}} \sum_{i<j} \sigma^z_i \sigma^z_j  \right.\\
        &+\left. \frac{\ii s \e^{\ii 2t} \sin^3\frac{t}{2}\cos\frac{t}{2} + \ii \sin^3\frac{t}{2}\cos^5\frac{t}{2} + \ii \sin^5\frac{t}{2}\cos^3\frac{t}{2}}{\left(s \e^{\ii 2t} + \cos^4\frac{t}{2}\right)^2 -  \sin^8\frac{t}{2} } \sum_{i<j<k} \sigma^z_i \sigma^z_j \sigma^z_k\right).
    \end{aligned}
    \label{eq:realistic}
\end{equation}
\end{widetext}
When $t=\pi$, one may check that the above expression reduces to the correct projection $(I+sB_{p_0})/2$. When $t\neq \pi$, $M_s$ stands for an imperfect measurement operator.

To understand the effect of the realistic error model Eq. \eqref{eq:realistic}, we compare it with the simplified model Eq. \eqref{eq:imperfect-measurement}. 
As a picture of how decoding fails for imperfect initial states, recall that 
those states contain superpositions of defects (stabilizer generator = -1 components) whose amplitudes do not decay with the distance between defects. The consequence is that the failure probability of decoding will not decay with code distance.
We notice that in its expression the first factor $ 1 +s \e^{-\ii 2t} \cos^4(t/2) + s \e^{-\ii 2t} \sin^4(t/2) B_{p_0}$ is similar to the imperfect measurement operator discussed in Eq. \eqref{eq:imperfect-measurement} $\exp({\beta s_{p_0} B_{p_0}/2 }) = \cosh (\beta/2) + s_{p_0} \sinh(\beta/2) B_{p_0}$. 
{Recall that in the simple model depicted in Fig.~\ref{fig:toric}(c), the absence of a finite measurement error threshold is attributed to the superposition of defects with $B_{p_0}=-1$. The amplitudes of these defects remain constant regardless of their spatial separation and depend solely on their quantity. In contrast, the realistic model illustrated in Fig.~\ref{fig:toric}(b) incorporates the term $\exp({\beta s_{p_0} B_{p_0}/2})$. This inclusion leads to superpositions analogous to those in the simple model, thereby suggesting that the issue of lacking a finite threshold persists in the realistic scenario.
Furthermore, there is an additional factor, which can be viewed as coherent errors appearing on data qubits. These coherent will further lead to $A_v = -1$ defects in the initial code state (see Sec. SVIII of SI).
So while discussing error correction, aside from the previously discussed consequences, coherent errors are likely to inflict greater damage on the error correction process, resulting in deteriorated performance. We also notice that under this realistic measurement model, if we define logical states by applying logical operators to the imperfect initial state, those states will not be orthogonal to each other, making it much harder to solve exactly. Finally, we stress that the preceding discussion concerning the realistic model relies on model-based calculations and partially qualitative reasoning. A rigorous mathematical proof for general cases is essential and merits comprehensive investigation in future studies.}

%When $t=\pi$, one may check that the above expression reduces to the correct projection $(I+sB_{p_0})/2$. When $t\neq \pi$, $M_s$ stands for an imperfect measurement operator. We notice that in its expression the first factor $ 1 +s \e^{-\ii 2t} \cos^4(t/2) + s \e^{-\ii 2t} \sin^4(t/2) B_{p_0}$ is similar to the imperfect measurement operator discussed in Eq. \eqref{eq:imperfect-measurement} $\exp({\beta s_{p_0} B_{p_0}/2 }) = \cosh (\beta/2) + s_{p_0} \sinh(\beta/2) B_{p_0}$. However, there is an additional factor, which can be viewed as coherent errors appearing on data qubits. So while discussing error correction, aside from the consequence talked about before, coherent errors will cause more problems in the QEC performance.

%It is evident that the employment of either a single-round measurement preparation or a multi-round preparation targeting general logical states is expected to yield states that are undecodable. 

For a small finite code, when considering the application of the multi-round measurement preparation protocol to specific logical states within the Pauli basis, its effectiveness is brought into question for more realistic measurement circuits.
In this scenario, the implementation of the $A_{v_0}$ measurement is necessitated for the correction of the $Z$ coherent error strings induced by the $B_{p_0}$ measurement. Given that these two distinct types of imperfect stabilizer measurement operators are non-commutative, their resulting measurement outcomes are inherently interdependent.
%In this case, $A_{v_0}$ measurement is required to correct the $Z$ coherent error strings induced by $B_{p_0}$ measurement. The two types of imperfect stabilizer measurement operator do not commute with each other, so their measurement outcomes will be interrelated. 
%The efficacy of the multi-round state preparation protocol in reducing residual logical errors, or its functionality in subsequent error correction cycles, remains uncertain here. 
Future investigations could benefit from exploring a hybrid decoding approach that combines both $A_{v_0}$ and $B_{p_0}$ stabilizer measurement data.

%We also notice that under this realistic measurement model, if we define logical states by applying logical operators to the imperfect initial state, those states will not be orthogonal to each other, making it much harder to analyze theoretically. 

\section{Discussions}
\label{sec:conclusion}
We study the performance of toric code QEC under the influence of various important error types, including imperfect measurement circuits for both preparation and error detection, as well as the qubit-level stochastic Pauli noise.
We establish a connection between the corresponding QEC performance and a novel statistical mechanical model. 
%Our result demonstrates that toric code prepared by a single layer of stabilizer measurement are unable to shield quantum information from local Pauli perturbations, and at least super-logarithm code distance rounds of measurements are required for preparing a QEC utilizable state. 
Our statistical mechanical analysis reveals the mechanism of how the imperfect initial code states affect the following error correction cycle.
%our result demonstrates how imperfect measurement-prepared toric code states are unable to shield quantum information from local Pauli perturbations.
We emphasize that the code state preparation requires special attention, which was rarely discussed in previous theoretical studies of QEC error thresholds. A naive preparation protocol could destroy the ability to shield quantum information from local Pauli perturbations.
%Although our main analytical investigation focuses on a simplified model (Fig. \ref{fig:toric}(c)), coherent noises in measurement preparation should nevertheless cause the same kind of damage in more general situations. 
%For example, in a more realistic measurement circuit (Fig. \ref{fig:toric}(b)), coherent noises on entanglement gates still lead to timelike non-local measurement error strings in decoding and hence have qualitatively the same effect.
%Furthermore, they also result in additional coherent error strings on the data qubits, which could worsen the performance. 
%In the simplified error model, the imperfect preparation problem could be amended by $\omega(\log d)$ rounds of preparation measurements. However, in the more realistic setups with a circuit error model \eqref{eq:realistic} and both $X$ and $Z$ stochastic Pauli errors, the situation is more complicated due to the accumulation of coherent error strings and correlation between imperfect $B_p$ and $A_v$ measurement, where actual performance there is still an open question.
Although focusing on toric code in the calculations, we expect our results hold for general boundary conditions of large enough surface codes since they are not expect to affect the qualitative properties of the SM model in the thermodynamic limit. 
In addition, we mention that QEC code states could be prepared by pure unitaries instead of measurements \cite{LinkeFT2017,takitaExperimentalDemonstrationFaultTolerant2017,ZhaoPRL-22}. However in large systems, the typical circuit depth of unitary preparation is larger than measurement preparation in the ideal case, and its resilience against coherent deviations could be even worse.

Before closure,  it is important to highlight that our results indicate a different impact based on the scale of quantum codes. Specifically, smaller quantum codes, typically situated within the $1/\log d$ threshold, still exhibit resilience against the outlined challenges (Sec.SVII of SI~\cite{SupMat}), thereby maintaining robust QEC capabilities.
Current small-scale experiments of surface code QEC
\cite{Erhard21Nature,Andersen20NP,GoogleAI21Nature,Marques22NP,ZhaoPRL-22,bluvstein_logical_2024} are believed to stay in the effectively correctable region such that they provides positive results.
Conversely, our findings hold considerable significance for larger-scale systems, which are central to the realization of application-level fault-tolerant quantum computing. As the scale of the system expands, the criticality of code space preparation must be significantly intensified, accompanied by an increased consumption of resources. This observation emphasizes the crucial need for a focus on refining state preparation protocols, a measure essential to safeguarding the efficacy and feasibility of QEC for advanced fault-tolerant quantum computations.

\bigskip
\noindent
\textbf{{Acknowledgments}}

Authors thank Guo-Yi Zhu for the discussions on the finite-size effect of the SM model, and Kenneth Brown for pointing out the state preparation with the multiple-round measurement protocols. We thank Jing-Yuan Chen, Ying-Fei Gu, Xie Chen, Guanyu Zhu, Ying Li, Li Rao, and Qinghong Yang for helpful discussions.
This work is supported by National Natural Science Foundation of China (Grants No. 92365111), the Innovation Program for Quantum Science and Technology (Grant No.~2021ZD0302400) and Beijing Natural Science Foundation (Grant No. Z220002).

\bigskip
\noindent
\textbf{{Competing interests}}

\noindent
The authors declare no competing interests.

\bigskip
\noindent
\textbf{{Data availability}}

\noindent
Data sharing is not applicable to this article, as no datasets were generated or analyzed during the current study.

%\bibliographystyle{apsrev4-1} 
%\bibliography{ref.bib}

%

\onecolumngrid
\newpage
\includegraphics[page=1, width=\textwidth, height=\textheight,trim=2cm 2cm 2cm 2cm, clip]{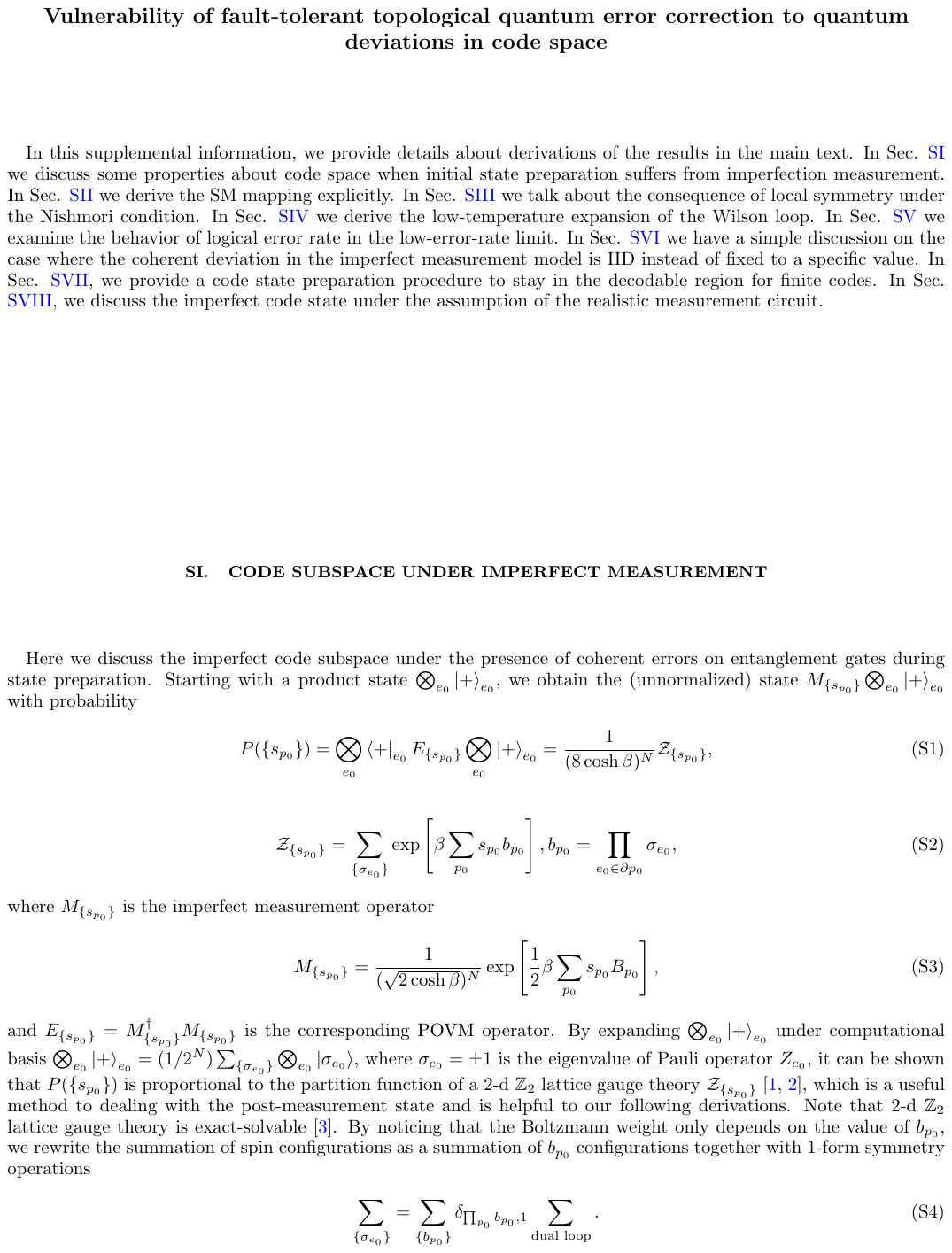}
\includegraphics[page=2, width=\textwidth, height=\textheight,trim=2cm 2cm 2cm 2cm, clip]{SI-FTQC-V3.pdf}
\includegraphics[page=3, width=\textwidth, height=\textheight,trim=2cm 2cm 2cm 2cm, clip]{SI-FTQC-V3.pdf}
\includegraphics[page=4, width=\textwidth, height=\textheight,trim=2cm 2cm 2cm 2cm, clip]{SI-FTQC-V3.pdf}
\includegraphics[page=5, width=\textwidth, height=\textheight,trim=2cm 2cm 2cm 2cm, clip]{SI-FTQC-V3.pdf}
\includegraphics[page=6, width=\textwidth, height=\textheight,trim=2cm 2cm 2cm 2cm, clip]{SI-FTQC-V3.pdf}
\includegraphics[page=7, width=\textwidth, height=\textheight,trim=2cm 2cm 2cm 2cm, clip]{SI-FTQC-V3.pdf}
\includegraphics[page=8, width=\textwidth, height=\textheight,trim=2cm 2cm 2cm 2cm, clip]{SI-FTQC-V3.pdf}
\includegraphics[page=9, width=\textwidth, height=\textheight,trim=2cm 2cm 2cm 2cm, clip]{SI-FTQC-V3.pdf}
\includegraphics[page=10, width=\textwidth, height=\textheight,trim=2cm 2cm 2cm 2cm, clip]{SI-FTQC-V3.pdf}
\includegraphics[page=11, width=\textwidth, height=\textheight,trim=2cm 2cm 2cm 2cm, clip]{SI-FTQC-V3.pdf}
\includegraphics[page=12, width=\textwidth, height=\textheight,trim=2cm 2cm 2cm 2cm, clip]{SI-FTQC-V3.pdf}
\includegraphics[page=13, width=\textwidth, height=\textheight,trim=2cm 2cm 2cm 2cm, clip]{SI-FTQC-V3.pdf}
\includegraphics[page=14, width=\textwidth, height=\textheight,trim=2cm 2cm 2cm 2cm, clip]{SI-FTQC-V3.pdf}
\includegraphics[page=15, width=\textwidth, height=\textheight,trim=2cm 2cm 2cm 2cm, clip]{SI-FTQC-V3.pdf}
\includegraphics[page=16, width=\textwidth, height=\textheight,trim=2cm 2cm 2cm 2cm, clip]{SI-FTQC-V3.pdf}
\includegraphics[page=17, width=\textwidth, height=\textheight,trim=2cm 2cm 2cm 2cm, clip]{SI-FTQC-V3.pdf}
\includegraphics[page=18, width=\textwidth, height=\textheight,trim=2cm 2cm 2cm 2cm, clip]{SI-FTQC-V3.pdf}
\includegraphics[page=19, width=\textwidth, height=\textheight,trim=2cm 2cm 2cm 2cm, clip]{SI-FTQC-V3.pdf}
\includegraphics[page=20, width=\textwidth, height=\textheight,trim=2cm 2cm 2cm 2cm, clip]{SI-FTQC-V3.pdf}
\includegraphics[page=21, width=\textwidth, height=\textheight,trim=2cm 2cm 2cm 2cm, clip]{SI-FTQC-V3.pdf}

%\includepdf[pages=1]{SI-FTQC-V3.pdf}

\end{document}